\documentclass[fleqn,usenatbib]{mnras}

\usepackage{newtxtext,newtxmath}
 
\usepackage[T1]{fontenc}

\DeclareRobustCommand{\VAN}[3]{#2}
\let\VANthebibliography\thebibliography
\def\thebibliography{\DeclareRobustCommand{\VAN}[3]{##3}\VANthebibliography}

\usepackage{graphicx}	
\usepackage{amsmath}	
\usepackage{amssymb}	

\title[Age-Divided Populations from Full Spectrum Fitting]{Age-Divided Mean Stellar Populations from Full Spectrum Fitting as the Simplified Star Formation and Chemical Evolution History of a Galaxy: Methodology and Reliability}

\author[J. H. Lee et al.]{
Joon Hyeop Lee$^{1}$\thanks{E-mail: jhl@kasi.re.kr},
Mina Pak$^{1, 2, 5}$,
Hyunjin Jeong$^{1}$,
Sree Oh$^{3, 4, 5}$
\\
$^{1}$Korea Astronomy and Space Science Institute, Daejeon 34055, Republic of Korea \\
$^{2}$School of Mathematical and Physical Sciences, Macquarie University, Sydney, NSW 2109, Australia \\
$^{3}$Department of Astronomy and Yonsei University Observatory, Yonsei University, Seoul 03722, Republic of Korea \\
$^{4}$Research School of Astronomy and Astrophysics, Australian National University, Canberra, ACT 2611, Australia \\
$^{5}$ARC Centre of Excellence for All Sky Astropysics in 3 Dimensions (ASTRO 3D), Australia \\
}

\date{Accepted 2023 March 8. Received 2023 March 8; in original form 2022 October 13}

\pubyear{2023}

\begin{document}
\label{firstpage}
\pagerange{\pageref{firstpage}--\pageref{lastpage}}
\maketitle

\begin{abstract}
We introduce a practical methodology for investigating the star formation and chemical evolution history of a galaxy: age-divided mean stellar populations (ADPs) from full spectrum fitting. In this method, the mass-weighted mean stellar populations and mass fractions ($f_{\textrm{mass}}$) of young and old stellar components in a galaxy are separately estimated, which are divided with an age cut (selected to be $10^{9.5}$ yr $\approx3.2$ Gyr in this paper). To examine the statistical reliability of ADPs, we generate 10,000 artificial galaxy spectra, each of which consists of five random simple stellar population components. Using the Penalized PiXel-Fitting (pPXF) package, we conduct full spectrum fitting to the artificial spectra with noise as a function of wavelength, imitating the real noise of Sydney-Australian Astronomical Observatory Multi-object Integral field spectrograph (SAMI) galaxies. As a result, the $\Delta$ (= output $-$ input) of age and metallicity appears to significantly depend on not only signal-to-noise ratio (S/N), but also luminosity fractions ($f_{\textrm{lum}}$) of young and old components. At given S/N and $f_{\textrm{lum}}$, $\Delta$ of young components tends to be larger than $\Delta$ of old components; e.g., $\sigma(\Delta$[M/H]$) \sim0.40$ versus 0.23 at S/N = 30 and $f_{\textrm{lum}}=50$ per cent. The age-metallicity degeneracy appears to be insignificant, but $\Delta$log(age/yr) shows an obvious correlation with $\Delta f_{\textrm{mass}}$ for young stellar components ($\mathcal{R}\sim0.6$).  The impact of dust attenuation and emission lines appears to be mostly insignificant. We discuss how this methodology can be applied to spectroscopic studies of the formation histories of galaxies, with a few examples of SAMI galaxies.
\end{abstract}




\section{INTRODUCTION}\label{intro}

Understanding the formation histories of galaxies is one of the major goals in extragalactic astronomy. Various approaches have been tried to achieve this goal, according to available data sets and more specific purposes: the analysis of star clusters in a galaxy \citep[e.g.,][]{dob10,for18a,for18b,ko18,vil19,dol21}, photometric profiling of individual galaxies \citep[e.g.,][]{hea15,kim15,her16,bou18,byu18,wat19,rom21}, tracing the properties of a bulk of galaxies as a function of cosmic time \citep[e.g.,][]{van10,alb16,fos17,mun17,mow19,chu21}, and so on. Such approaches often rely on statistics, which mainly look for a common history of galaxies in a similar type. A large sample of galaxies spanning over a wide range of redshift is particularly useful for such studies, providing a chronological view on galaxy evolution as cosmic time flows. On the other hand, some methods can be used for finding out the individual history of each galaxy; for example, full spectrum fitting.

When we study stellar populations of a galaxy with its optical spectrum, two major methods are frequently used: absorption line strength fitting \citep[e.g.,][]{pro04,sch07,tho11,mcd15,sco17,pak19,pak21,kim20,bor22} and full spectrum fitting \citep[e.g.,][]{gon14,mcd15,wil15,che16,gue16,iye17,bar18,car18,bar20,ge18,han19,pet19,san19,joh21,vau22,wan22}. The absorption line strength fitting directly measures hydrogen and metal lines in stellar atmospheres. Thus, it may better estimate luminosity-weighted stellar age, metallicity and $\alpha$-element abundance, which are more sensitive to the properties of young and bright stars. On the other hand, the full spectrum fitting considers not only absorption lines but also the shape of continuum, and thus better reflects old stars, compared to the absorption line fitting. Moreover, it is possible to estimate mass-weighted mean stellar populations through the full spectrum fitting. Both of these approaches have their own merits, but the full spectrum fitting has an additional advantage that it constrains not only the mean stellar populations, but also the full history of star formation and chemical evolution of a galaxy \citep[e.g.,][]{toj07,nor15,she15,gon17,zho22}.

Conceptually, the estimation of the star formation and chemical evolution history through the full spectrum fitting is simple: when the observed spectrum is decomposed into several templates of simple stellar populations, we can reconstruct its formation history by ordering the stellar populations (their ages, metal abundances and mass fractions) along lookback time. 
Note that this reconstructed history may be the combined history of in-situ and ex-situ star formation, because usual full spectrum fitting by itself cannot separate those two kinds and such separation requires a different approach \citep[e.g.,][]{poc19}. Thus, the `star formation and chemical evolution history' in this paper does not distinguish mass assembly from star formation history.

The problem of the reconstructed history through full spectrum fitting is that it is not easy to know how reliable the decomposed stellar populations are. For this reason, until now, the full spectrum fitting has been used mostly for estimating the total mean stellar populations rather than the detailed history of galaxy formation. Even in the studies on the latter topic, the main interest has been mostly the statistically summed (not individual) star formation history, and moreover, the chemical evolution history was more rarely inspected through the full spectrum fitting. However, if it is possible to estimate the quantitative uncertainty of the star formation and chemical evolution history of an individual galaxy, even though it is in a simplified scheme, the full spectrum fitting would be much better utilised for understanding galaxy formation and evolution.

In this paper, we introduce a practical methodology for investigating the star formation and chemical evolution history of a galaxy: age-divided mean stellar populations (ADPs) from full spectrum fitting. In this method, the mass-weighted mean stellar populations (age and metallicity) and mass fractions ($f_{\textrm{mass}}$) of young and old stellar components in a galaxy are separately estimated. The goals of this paper are: (1) to estimate the statistical uncertainty of total mean stellar populations and ADPs from full spectrum fitting, and (2) to provide the uncertainty tables that can be practically applied to a specific combination of a data set and a full spectrum fitting tool: the Sydney-Australian Astronomical Observatory Multi-object Integral field spectrograph \citep[SAMI;][]{cro12} Galaxy Survey \citep{bry15} data and the Penalized PiXel-Fitting \citep[pPXF;][]{cap04,cap17,cap22} package. 
This paper is organised as follows. Section~\ref{data} introduces the data set (SAMI) and the analysis tool (pPXF) used in this work, including our pPXF running processes. Section~\ref{method} describes how to build the artificial input spectra and how to compare the input with the output from the pPXF fitting using ADPs. In section~\ref{result}, we check the possible degeneracy between stellar age, metallicity and mass fraction, as well as we present the dependence of the output $-$ input difference of the parameters (hereafter, $\Delta$) on signal-to-noise ratio (S/N) and luminosity fraction ($f_{\textrm{lum}}$). Section~\ref{discuss} discusses the results and shows how our results can be applied to the studies using actual spectroscopic data, with a few examples of SAMI galaxies. Finally, the paper is concluded in section~\ref{conclude}.

\section{DATA AND ANALYSIS}\label{data}

This paper partially utilises the SAMI Galaxy Survey Data Release Three \citep{cro21}, which provides spatially-resolved spectra of 3,068 galaxies at a redshift range of $z < 0.1$, observed with the SAMI instrument \citep{cro12} mounted on the 3.9-m Anglo-Australian Telescope at Siding Spring Observatory in Australia. 
The SAMI instrument has 13 hexabundles, each of which comprises 61 fibres with a 75 per cent
filling factor. The field of view of each hexabundle has 15 arcsec diameter, with 1.6-arcsec diameter of each fibre core. The spectroscopy was done with a blue arm (3700 -- 5700 {\AA}; R = 1730; 1.03 {\AA} per pixel) and a red arm (6250 -- 7350 {\AA}; R = 4500; 0.57 {\AA} per pixel) at the same time, which results in a total wavelength coverage of 3700 -- 7350 {\AA} with a small gap between the blue and red arms. The full width at half maximum (FWHM) is 2.65 {\AA} and 1.61 {\AA} for the blue and red data cubes, respectively.
For the full spectrum fitting, the blue and red spectra are merged into a single spectrum by smoothing and resampling the red spectrum so that it has the same FWHM and resolution as the blue spectrum.
The SAMI Galaxy Survey covers eight galaxy clusters as well as three regions of the Galaxy And Mass Assembly \citep[GAMA;][]{dri11} survey. The details of the SAMI galaxy sample are described in \citet{bry15} and \citet{owe17}.

We use the pPXF package \citep{cap04,cap17} for the full spectrum fitting, with the stellar population models from the Medium resolution INT Library of Empirical Spectra \citep[MILES;][]{vaz10}. Among the various options for the simple stellar population models, we adopt the Padova+00 isochrone \citep{gir00} with the initial mass function with a unimodal logarithmic slope $\Gamma = 1.3$. The template ages and metallicities we used are: age [Gyr] = [0.063, 0.079, 0.10, 0.13, 0.16, 0.20, 0.25, 0.32, 0.40, 0.50, 0.63, 0.79, 1.00, 1.26, 1.58, 2.00, 2.51, 3.16, 3.98, 5.01, 6.31, 7.94, 10.00, 12.59, 15.85] and [M/H] = [$-1.71$, $-1.31$, $-0.71$, $-0.40$, 0.00, $+0.22$].

The processes for running pPXF are summarised as follows:
\begin{itemize}
\item[(1)] First run with rough (randomly selected) initial guesses for velocity, velocity dispersion and noise.
\item[(2)] Second run with more realistic initial guesses based on the result of the first run.
\item[(3)] Third run with the bad pixel masks from the second run, with setting kinematics as free parameters (like in steps (1) and (2)) and with the degree of the additive Legendre polynomial to correct the template continuum (\emph{degree}) set to be 10. If dust reddening is considered, this is the final run.
\item[(4)] Fourth run if dust reddening is not considered, with fixing kinematics as obtained from the third run and with \emph{degree} = $-1$ and the degree of the multiplicative Legendre polynomial to correct the continuum shape (\emph{mdegree}) set to be 10.
\end{itemize}

We conduct the processes for a given spectrum with various options: on and off respectively for gas emission, dust reddening, and regularisation (\emph{regul} = 5, if on). These processes return the full list of templates fitted to a given spectrum (age, metallicity, and mass fraction) as well as the total mean stellar population. In this paper, all stellar populations are mass-weighted mean populations.

\begin{figure}
\centering
\includegraphics[width=\columnwidth]{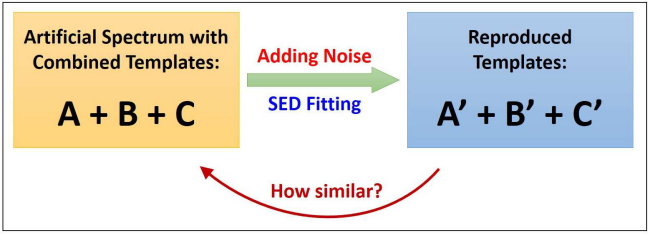}
\caption{Conceptual scheme of the reliability check of the templates reproduced by full spectrum fitting. \label{concept}}
\end{figure}

\begin{figure*}
\centering
\includegraphics[width=1.9\columnwidth]{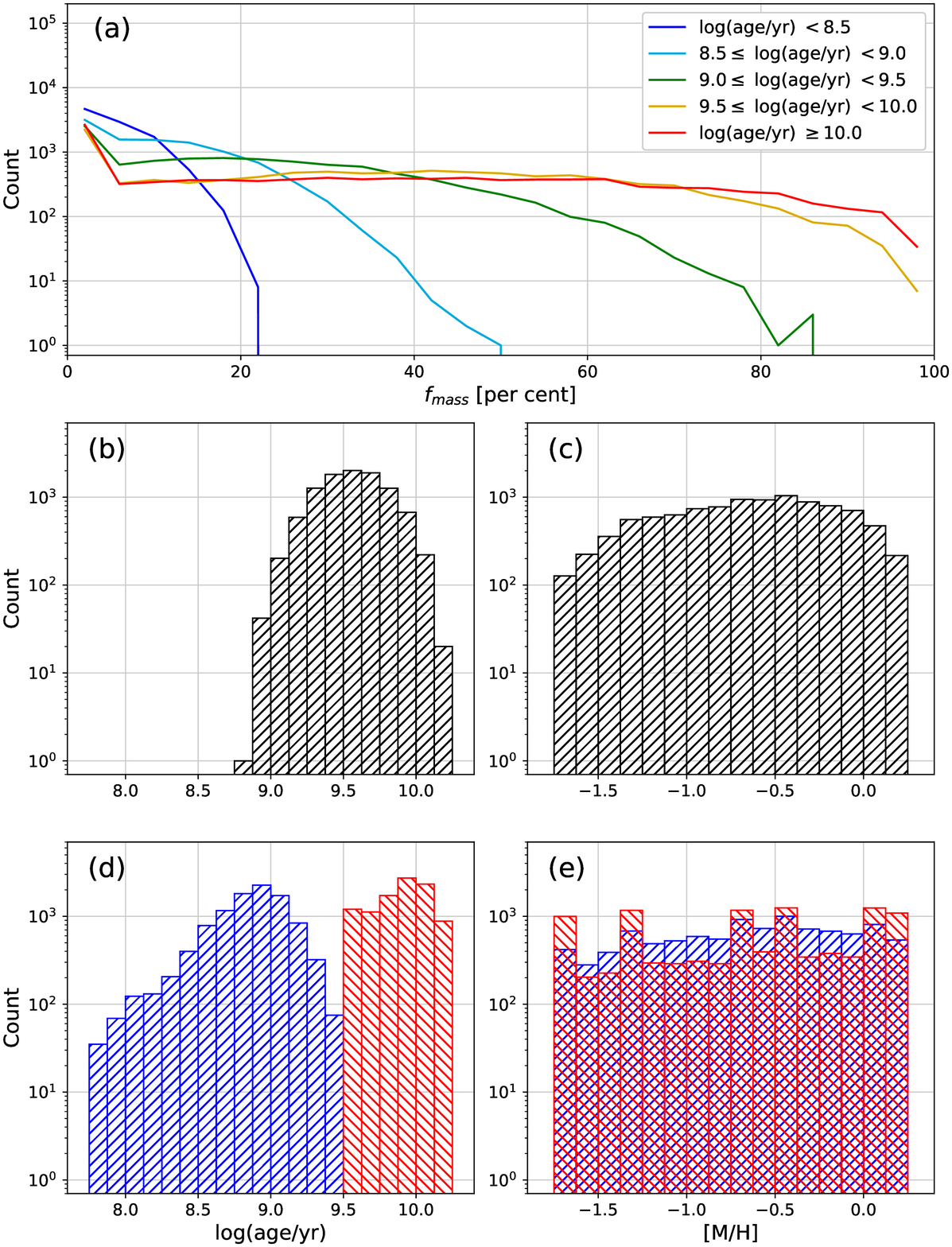}
\caption{The statistics for the 10,000 input spectra randomly produced. (a) The mass fraction ($f_{\textrm{mass}}$) distributions of the five age templates composing the input spectra. (b) Total mean age distribution of the input spectra. (c) Total mean metallicity distribution of the input spectra. (d) Mean age distributions, which are separately estimated for young ($< 10^{9.5} M_{\odot}$ yr; blue histogram) and old ($\geq 10^{9.5} M_{\odot}$ yr; red histogram) components of the input spectra. (e) Mean metallicity distributions, which are separately estimated for young and old components of the input spectra. \label{input}}
\end{figure*}

\section{METHODS}\label{method}

Figure~\ref{concept} summarises how to estimate the reliability of the star formation and chemical evolution history from full spectrum fitting. First, we build a set of artificial spectra by combining the stellar population model templates, which are randomly selected and thus represent various formation histories. To make these artificial spectra more realistic, we add noise as a function of wavelength, imitating observed spectra (in this paper, the SAMI spectra). Finally, we examine how well the pPXF results reproduce the input spectra, not only in the total mean stellar populations but also in ADPs.
Here, we describe each step in detail.

\subsection{Building Artificial Spectra}\label{buildspec}

We build a total of 10,000 artificial spectra with various star formation and chemical evolution histories. The building algorithms are as follows:
\begin{itemize}
\item[(1)] Each artificial spectrum is built by combining five simple stellar population model templates (from T1 to T5) among 25 ages $\times$ 6 metallicities = 150 templates from the MILES models. The five templates are selected from five different bins of age: T1 (log(age/yr) $< 8.5$), T2 ($8.5\leq$ log(age/yr) $<9.0$), T3 ($9.0\leq$ log(age/yr) $<9.5$), T4 ($9.5\leq$ log(age/yr) $<10.0$), and T5 (log(age/yr) $\geq 10.0$).
\item[(2)] Within each age bin, the selection of age and metallicity is fully random. For example, a T2 template may have any age and metallicity values among the MILES templates in $8.5\leq$ log(age/yr) $<9.0$ and $-1.71 \leq$ [M/H] $\leq +0.22$.
\item[(3)] The mass fraction ($f_{\textrm{mass}}$) of each selected template is determined to be a Gaussian random value, where the mean ($m$) and sigma ($\sigma$) of the Gaussian depend on the age bin: T1 ($m = 0.044$ and $\sigma = 0.050$), T2 ($m = 0.087$ and $\sigma = 0.100$), T3 ($m = 0.174$ and $\sigma = 0.200$), T4 ($m = 0.348$ and $\sigma = 0.400$), and T5 ($f_{\textrm{T5}}$ = 1 $-$ ($f_{\textrm{T1}}$ + $f_{\textrm{T2}}$ + $f_{\textrm{T3}}$ + $f_{\textrm{T4}}$)).
\end{itemize}

The Gaussian parameters were selected for the basic mass ratios to be $f_{\textrm{T1}}$:$f_{\textrm{T2}}$:$f_{\textrm{T3}}$:$f_{\textrm{T4}}$:$f_{\textrm{T5}}$ = 1:2:4:8:8, which roughly reflects the fact that known galaxies tend to have a large mass fraction of old populations and a small mass fraction of young populations. For example, \citet{van10} showed that 56 per cent of the stellar mass of a local massive galaxy at an environmental number density of $2\times 10^{-4}$ Mpc$^{-3}$ was already assembled at $z\sim 2$ ($\tau_{\textrm{lookback}} \sim 10$ Gyr). However, since the mass assembly history of a galaxy may significantly depend on its total mass and environment, we selected sufficiently large values as the Gaussian sigma. This procedure generates galaxy spectra with a variety of star formation and chemical evolution histories.

Figure~\ref{input} shows the distributions of several parameters describing the 10,000 random spectra. Extremely young (log(age/yr) < 8.5) components occupy a mass fraction up to $\sim$ 20 per cent, while extremely old (log(age/yr) $\geq 10.0$) components are distributed almost evenly from zero to a hundred per cent. The total mean ages of the random spectra range over $8.8 \lesssim$ log(age/yr) $\lesssim 10.3$, and their total mean metallicity is distributed over the entire range of the templates ($-1.71 \leq$ [M/H] $\leq +0.22$). When the population parameters are separately estimated for young (log(age/yr) $< 9.5$) and old (log(age/yr) $\geq 9.5$) components (i.e., ADPs), the age distribution covers almost the entire range of the templates ($7.8 \lesssim$ log(age/yr) $\lesssim 10.3$).
Note that the mean stellar age and metallicity in this paper are mass-weighted, and if luminosity-weighting were adopted, the age distribution would be more widely dispersed to younger ages making the overall distribution flatter.

\begin{figure}
\centering
\includegraphics[width=\columnwidth]{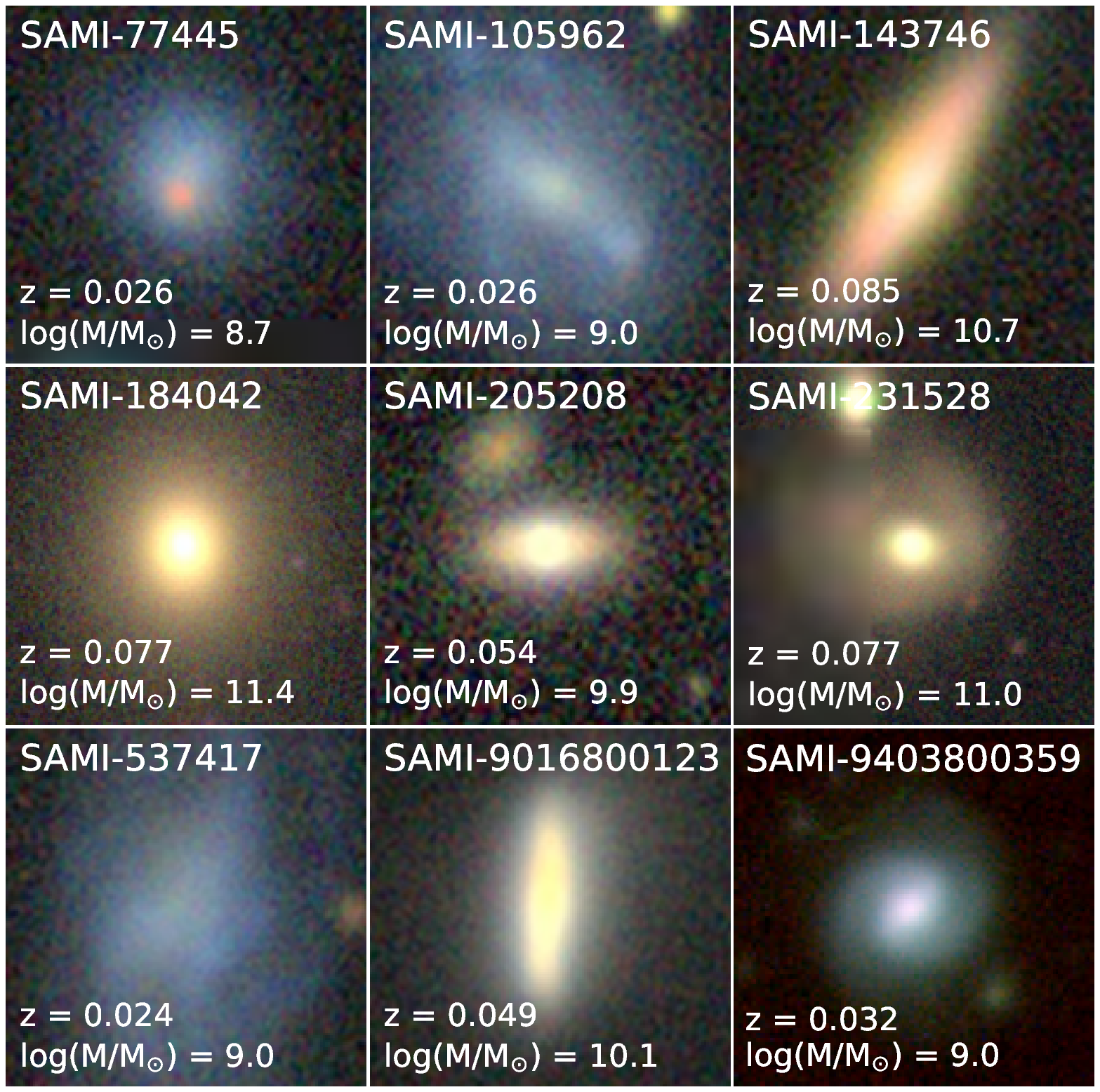}
\caption{Nine SAMI galaxies selected for the estimation of the noise function. The atlas images were retrieved from the Dark Energy Survey, and their redshifts and stellar masses are from the SAMI input catalogues. \label{nine}}
\end{figure}

\begin{figure}
\centering
\includegraphics[width=0.93\columnwidth]{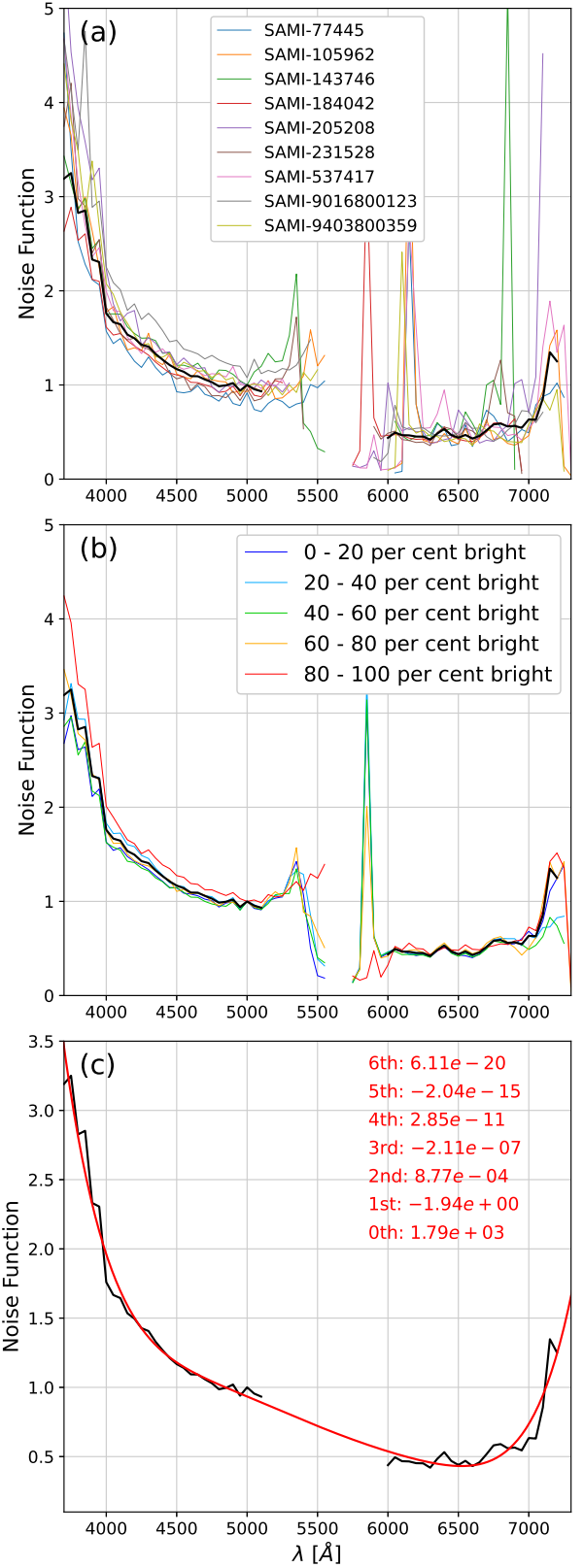}
\caption{Median noise functions, based on 436 spectra from the nine SAMI galaxies: (a) for each SAMI galaxy; (b) according to the brightness of each spectrum; and (c) the total median noise function (black line) and a sixth-order polynomial fit to it (red line). \label{noisef}}
\end{figure}

\begin{figure}
\centering
\includegraphics[width=0.9\columnwidth]{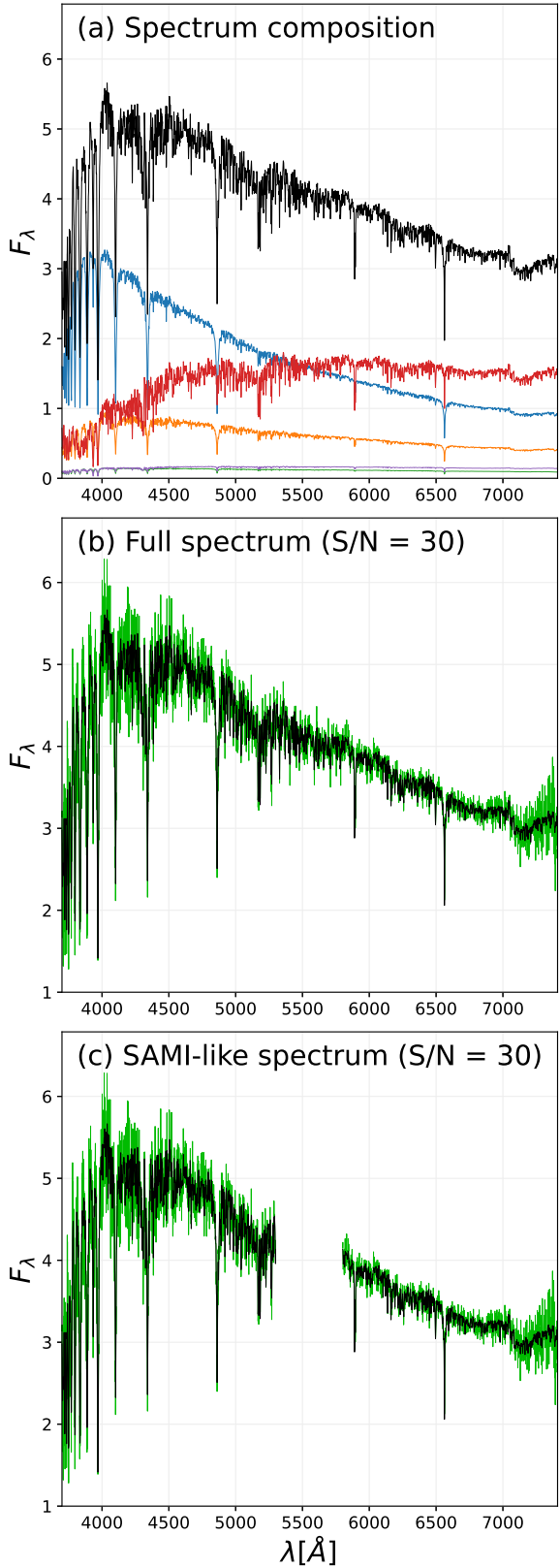}
\caption{An example of an artificially composed spectrum. (a) Five templates of simple stellar populations (coloured lines) are combined into a single spectrum (black line). (b) Random noise is added to the composite spectrum (S/N = 30, green line), based on the noise function fitted in Figure~\ref{noisef}. (c) The same as (b), but a SAMI-like wavelength gap is added (5300 -- 5800 \AA). \label{examsp}}
\end{figure}

\begin{figure}
\centering
\includegraphics[width=\columnwidth]{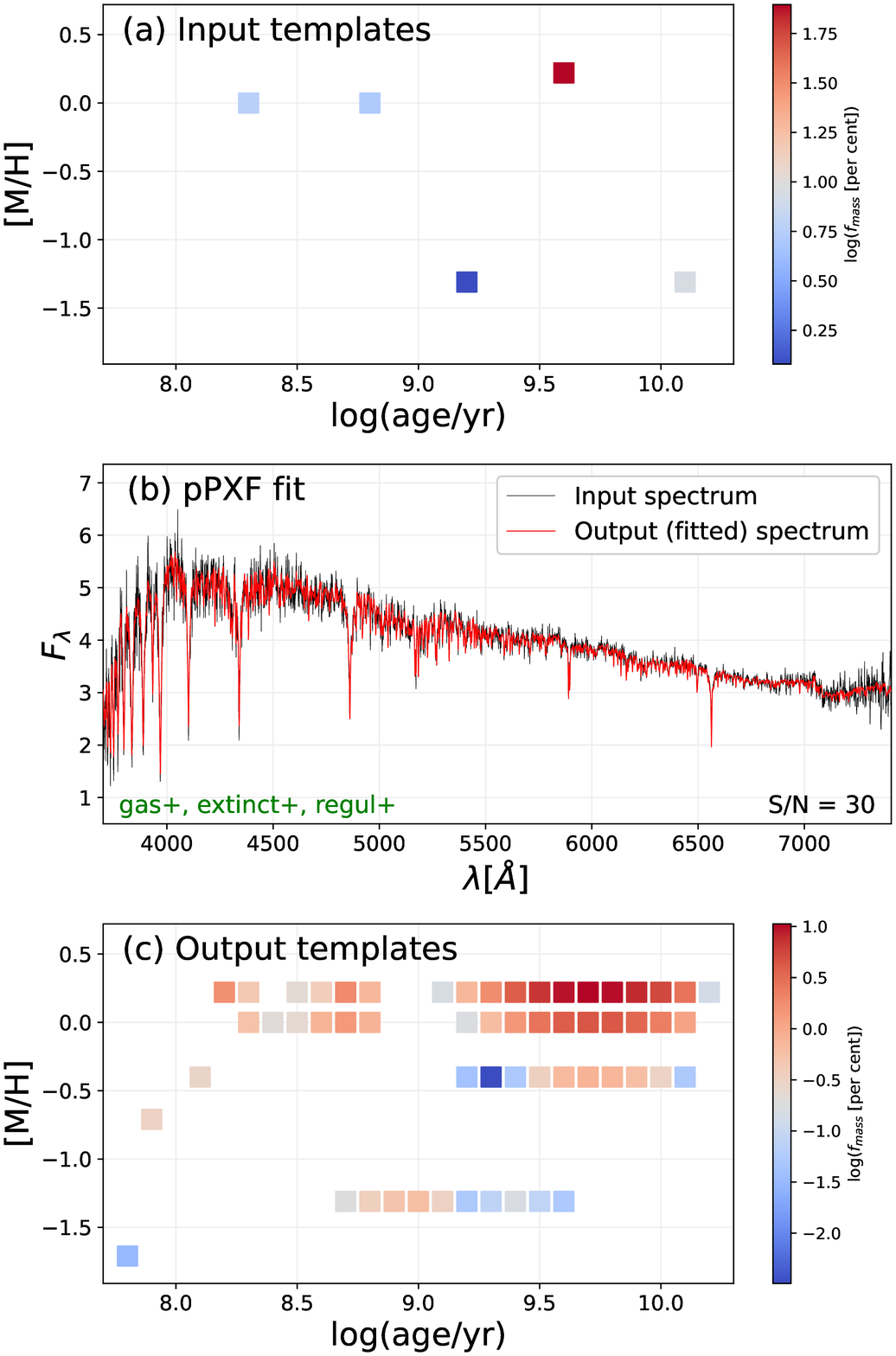}
\caption{Comparison between the input and output (pPXF-fitted) templates for an example spectrum (the same spectrum as Figure~\ref{examsp}). (a) Input templates on metallicity versus age diagram, with $f_{\textrm{mass}}$ of each template. (b) The pPXF fitting result, when gas emission, dust reddening, and regularisation options are all on. (c) Output templates from the pPXF fitting. \label{inout}}
\end{figure}

\subsection{Adding Noise}\label{addnoise}

Noise is a key element making a spectrum realistic, and at the same time, making it difficult to accurately estimate the stellar populations composing the observed spectrum. In reality, noise can arise due to various reasons, such as the intrinsic properties of the target, atmospheric conditions, instrument, and so on. Since it is not easy to reproduce such individual effects, here we use an empirical formula as a function of wavelength, based on the observational data from the SAMI Galaxy Survey. We selected nine SAMI galaxies that have various mass, morphology, colour and redshift, as shown in Figure~\ref{nine}. The atlas images are retrieved from the Dark Energy Survey\footnote{via https://www.legacysurvey.org} \citep[DES;][]{fla05}.
The number of spatially resolved spectra of the nine galaxies is 436, which are the results of Voronoi binning for their S/N to be larger than 10.

To determine the shape of noise as a function of wavelength, we devised a noise function $\mathcal{F_N}$, which is defined as:
\begin{equation}
\qquad \mathcal{F_N}(\lambda) = \cfrac{N(\lambda)}{\sqrt{F(\lambda)}} \quad ,
\end{equation}
where $N$ is noise, $F$ is flux, and $\lambda$ is wavelength.
The noise at a given wavelength ($N(\lambda')$) is estimated to be the standard deviation of the residual between the observed and fitted spectra inside a moving window of $\pm$10 {\AA}.
The noise function formula is based on the simple assumption that noise is determined by flux and wavelength only. Although such an assumption may not be true, practically it imitates the reality pretty well.
Since $\mathcal{F_N}$ is a quantity normalised by flux, we can focus on the noise dependence on wavelength, regardless of the source brightness that must be different among the nine SAMI galaxies and their individual bins.

Figure~\ref{noisef} shows the median noise functions (normalised to the values at $\lambda = 5000$ {\AA}) according to various subsamples, and a sixth order polynomial fit to the total median $\mathcal{F_N}$. Although there are small deviations between different galaxies and according to integrated brightnesses of individual spectra, the overall shapes of the noise functions appear to be consistent with one another. This noise function may slightly change according to the selection of sample galaxies, but obtaining the approximate shape of the noise function is enough for our purpose. However, since the noise function may significantly depend on observational data sets, our results from now on can be safely applied to the SAMI galaxies only.

An example of a random spectrum is presented in Figure~\ref{examsp}. The spectrum is composed of five MILES model templates, which becomes a more realistic one by adding noise that follows the noise function estimated in Figure~\ref{noisef}. 
In detail, we add noise to the artificial spectra as follows:
\begin{itemize}
\item[(1)] We determine Gaussian sigma values as a function of wavelength to be $\sigma(\lambda) = \mathcal{F_N}(\lambda) \times {\sqrt{F(\lambda)}}$, where $F(\lambda)$ is a model spectrum produced in Section~\ref{buildspec}.
\item[(2)] We build a noise spectrum from the random Gaussian generation at each wavelength bin, with Gaussian center = 0 and sigma $= \sigma(\lambda)$.
\item[(3)] We combine the model spectrum and the noise spectrum, from which we estimate the total median S/N ($\equiv$ (S/N)$_0$).
\item[(4)] When the aimed S/N is (S/N)$_{\textrm{A}}$, we build the final spectrum to be the model spectrum + the noise spectrum / $R$, where $R =$ (S/N)$_{\textrm{A}}$ / (S/N)$_0$.
\end{itemize} 
The SAMI-like version of the spectrum is made by adding a gap at an intermediate wavelength range (here, 5300 -- 5800 {\AA}). In the figure, the example spectrum has S/N = 30, but the S/N can be easily adjusted using the noise function.

\begin{figure*}
\centering
\includegraphics[width=2.0\columnwidth]{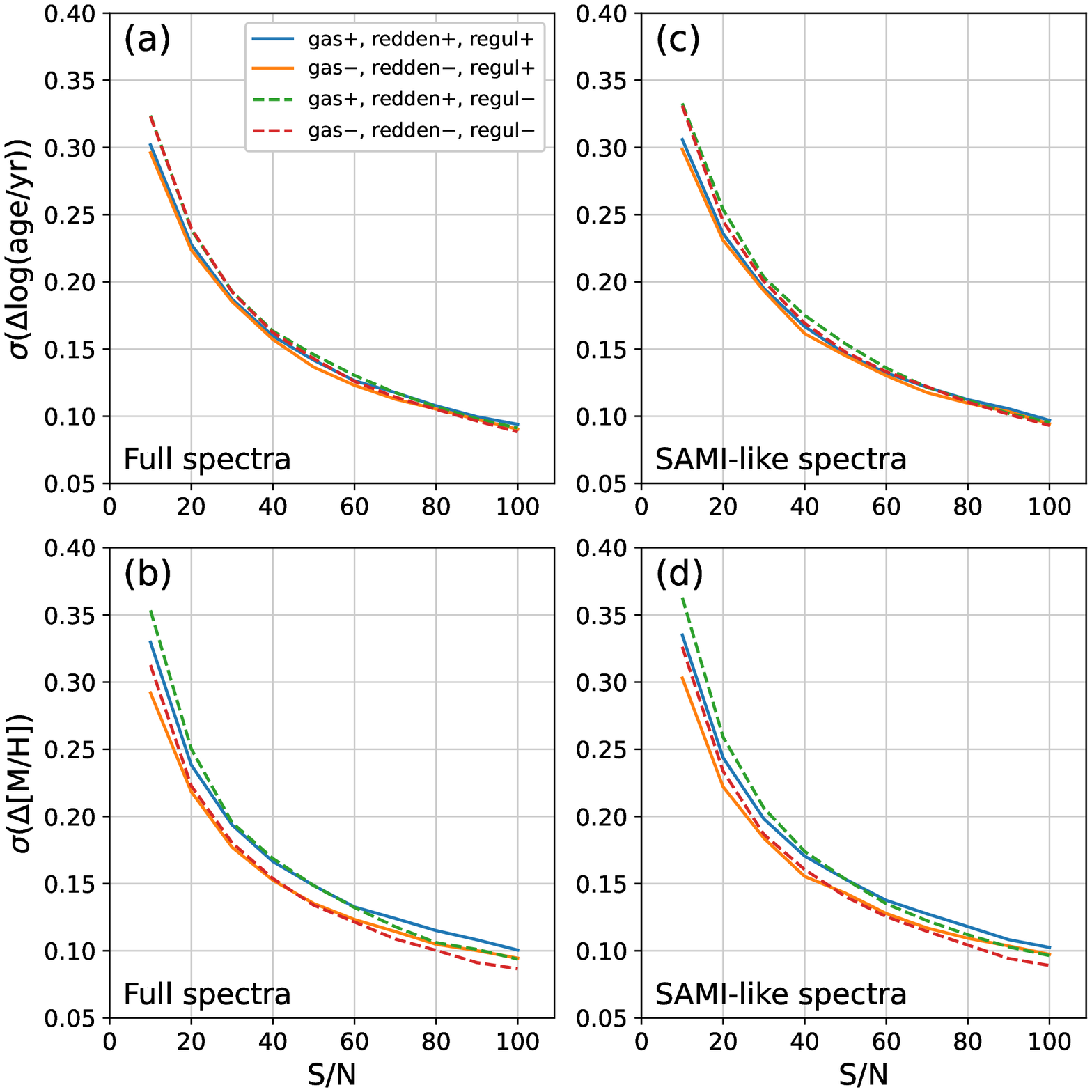}
\caption{(a) Standard deviation of $\Delta$log(age/yr) as a function of S/N, when each random spectrum is fully used. The line varies according to the various options of the pPXF fitting, but the differences are relatively small. (b) Standard deviation of $\Delta$[M/H] as a function of S/N, when each random spectrum is fully used. (c) Standard deviation of $\Delta$log(age/yr) as a function of S/N, when each spectrum is cut as a SAMI spectrum. (d) Standard deviation of $\Delta$[M/H] as a function of S/N, when each spectrum is cut as a SAMI spectrum. \label{singlesn}}
\end{figure*}

\begin{figure}
\centering
\includegraphics[width=0.82\columnwidth]{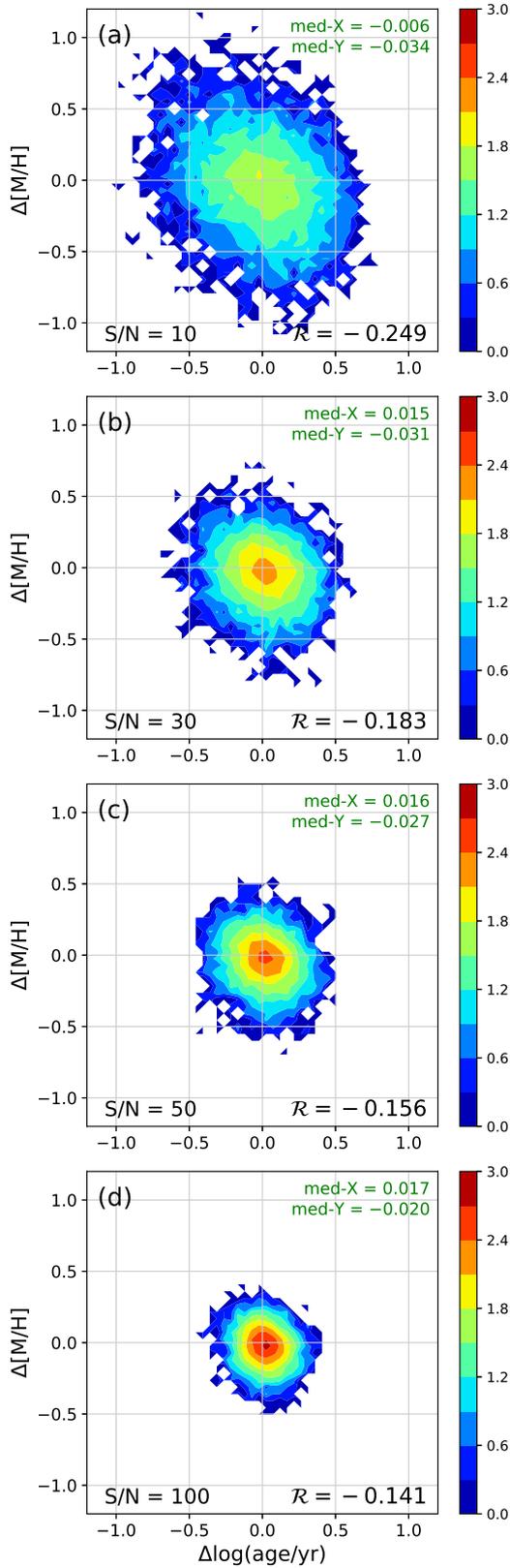}
\caption{Correlation between $\Delta$[M/H] and $\Delta$log(age/yr) of the total mean stellar populations, presented by contour maps of log(the number of data points), when (a) S/N = 10, (b) S/N = 30, (c) S/N = 50, and (d) S/N = 100. Pearson correlation coefficient ($\mathcal{R}$) is denoted in each panel. Each p-value is extremely small ($\ll 0.0001$). The median values of X- and Y-axis parameters (med-X and med-Y) are also denoted in each panel. \label{agemet}}
\end{figure}

\begin{figure}
\centering
\includegraphics[width=\columnwidth]{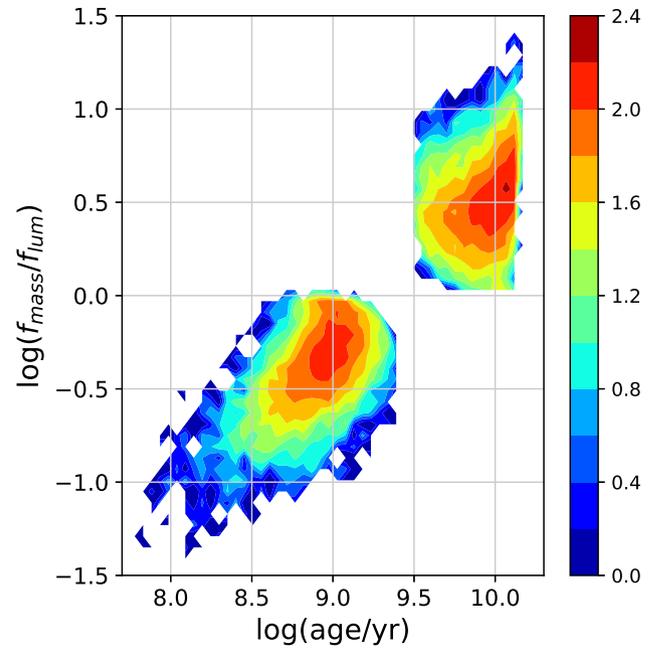}
\caption{The ratio between mass fraction and luminosity fraction of the age-divided mean stellar populations as a function of age, presented by a contour map of log(the number of data points). \label{masslum}}
\end{figure}

\begin{figure*}
\centering
\includegraphics[width=1.6\columnwidth]{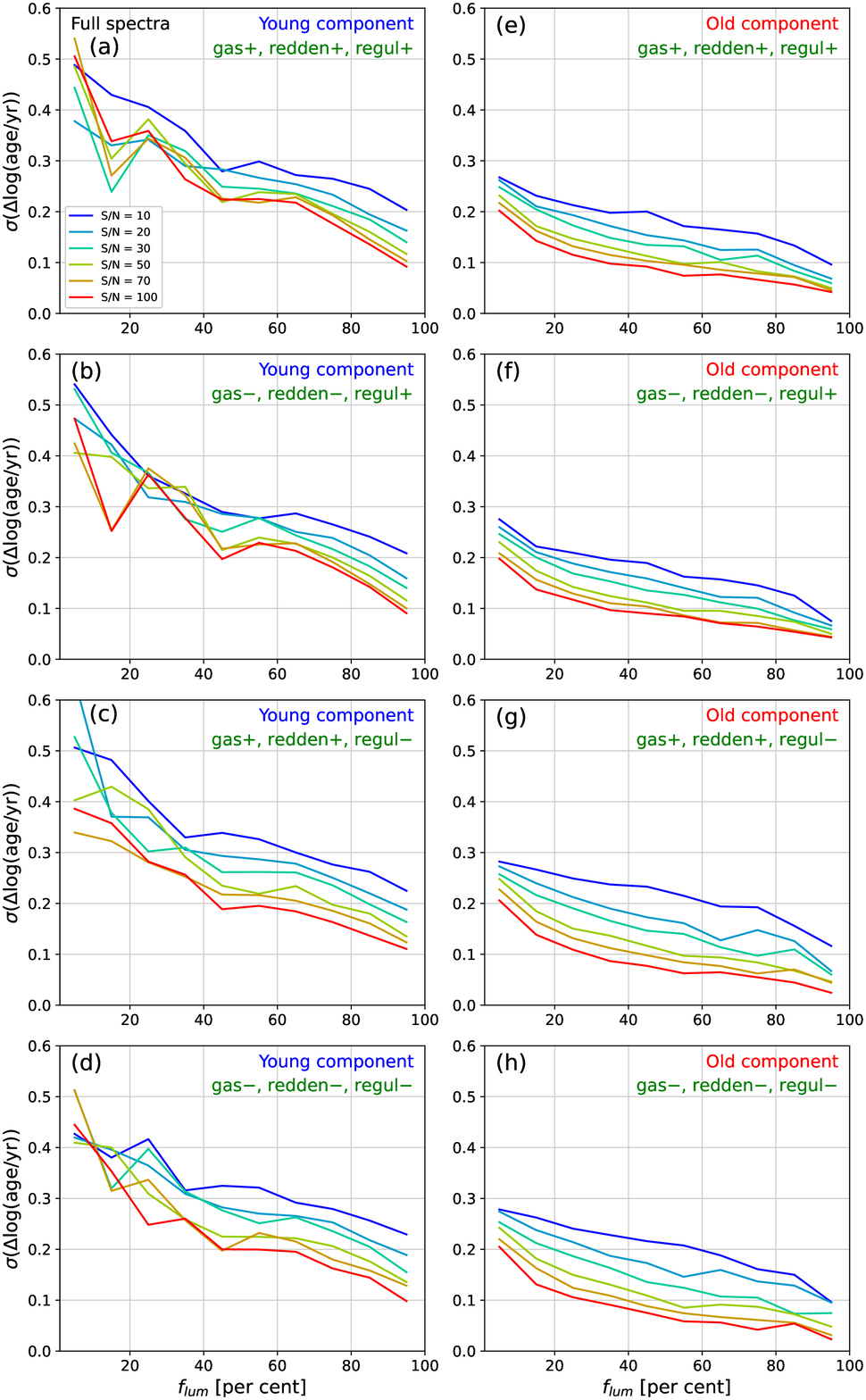}
\caption{Standard deviation of $\Delta$log(age/yr) as a function of luminosity fraction, when the pPXF fitting is done for the full spectra: (a) - (d) for young components, and (e) - (h) for old components, with various options of gas emission, dust reddening, and regularisation. \label{age1}}
\end{figure*}

\begin{figure*}
\centering
\includegraphics[width=1.6\columnwidth]{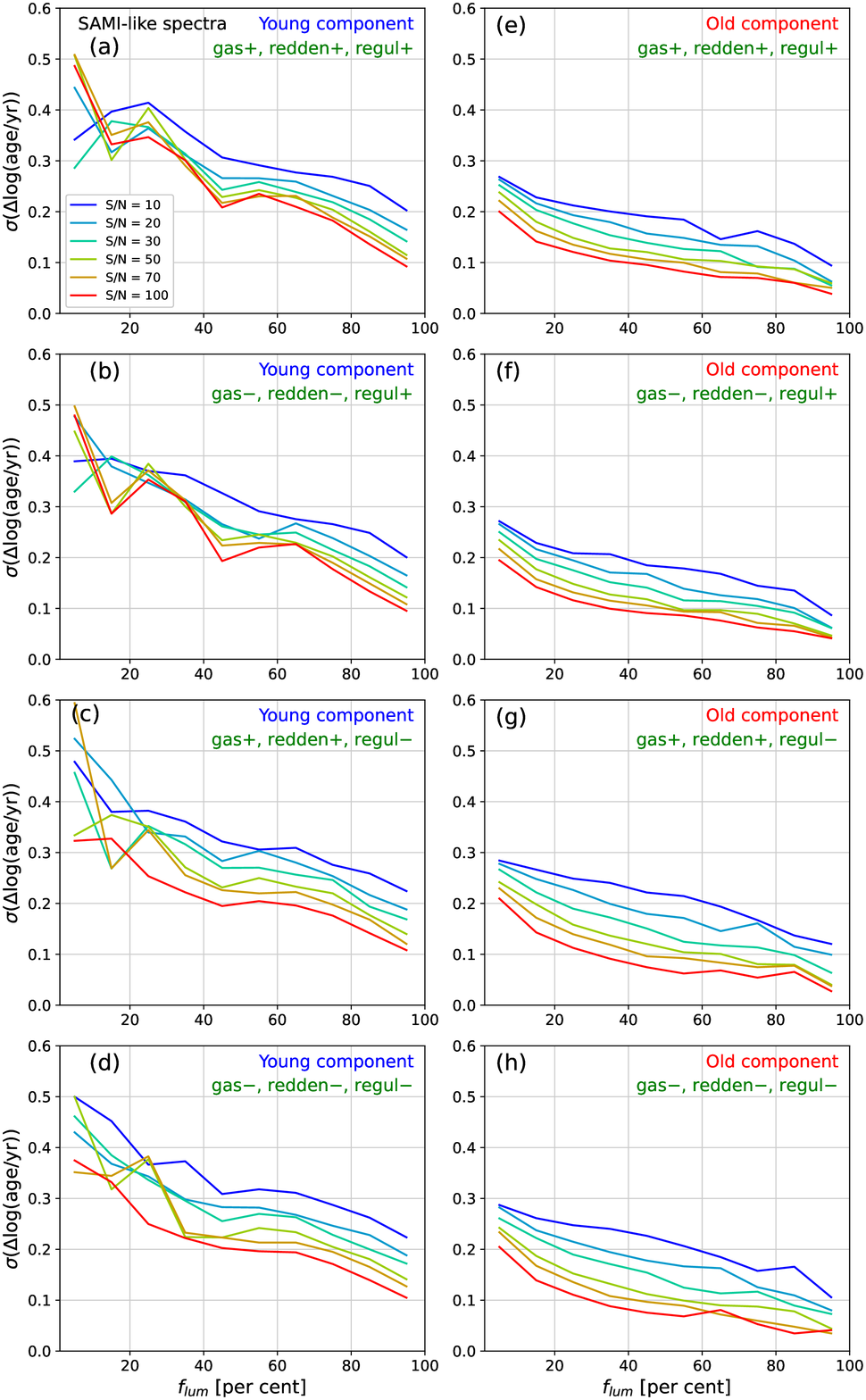}
\caption{The same as Figure~\ref{age1}, but when the pPXF fitting is done for the SAMI-like spectra. \label{age2}}
\end{figure*}

\begin{figure*}
\centering
\includegraphics[width=1.6\columnwidth]{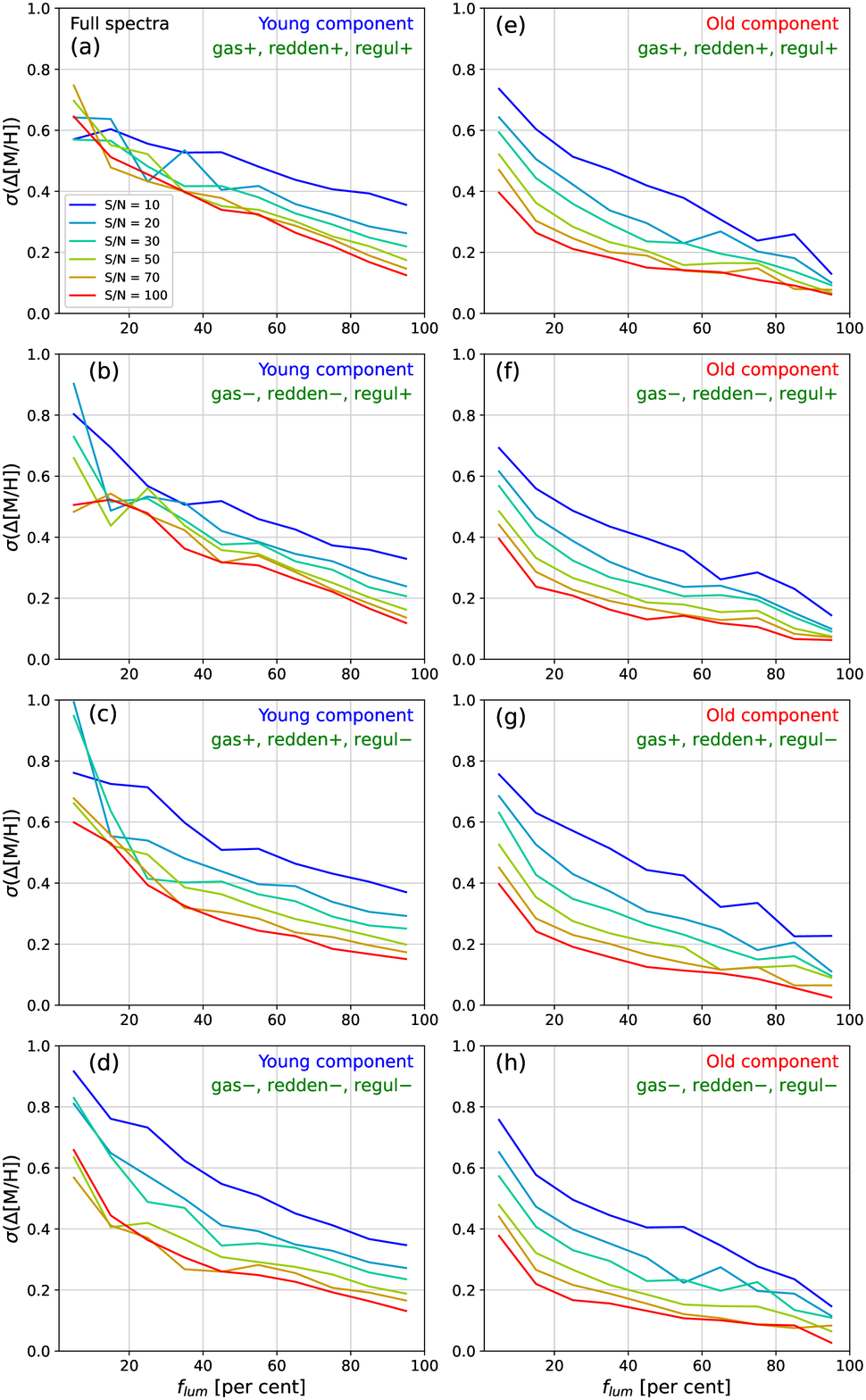}
\caption{Standard deviation of $\Delta$[M/H] as a function of luminosity fraction, when the pPXF fitting is done for the full spectra: (a) - (d) for young components, and (e) - (h) for old components, with various options of gas emission, dust reddening, and regularisation. \label{met1}}
\end{figure*}

\begin{figure*}
\centering
\includegraphics[width=1.6\columnwidth]{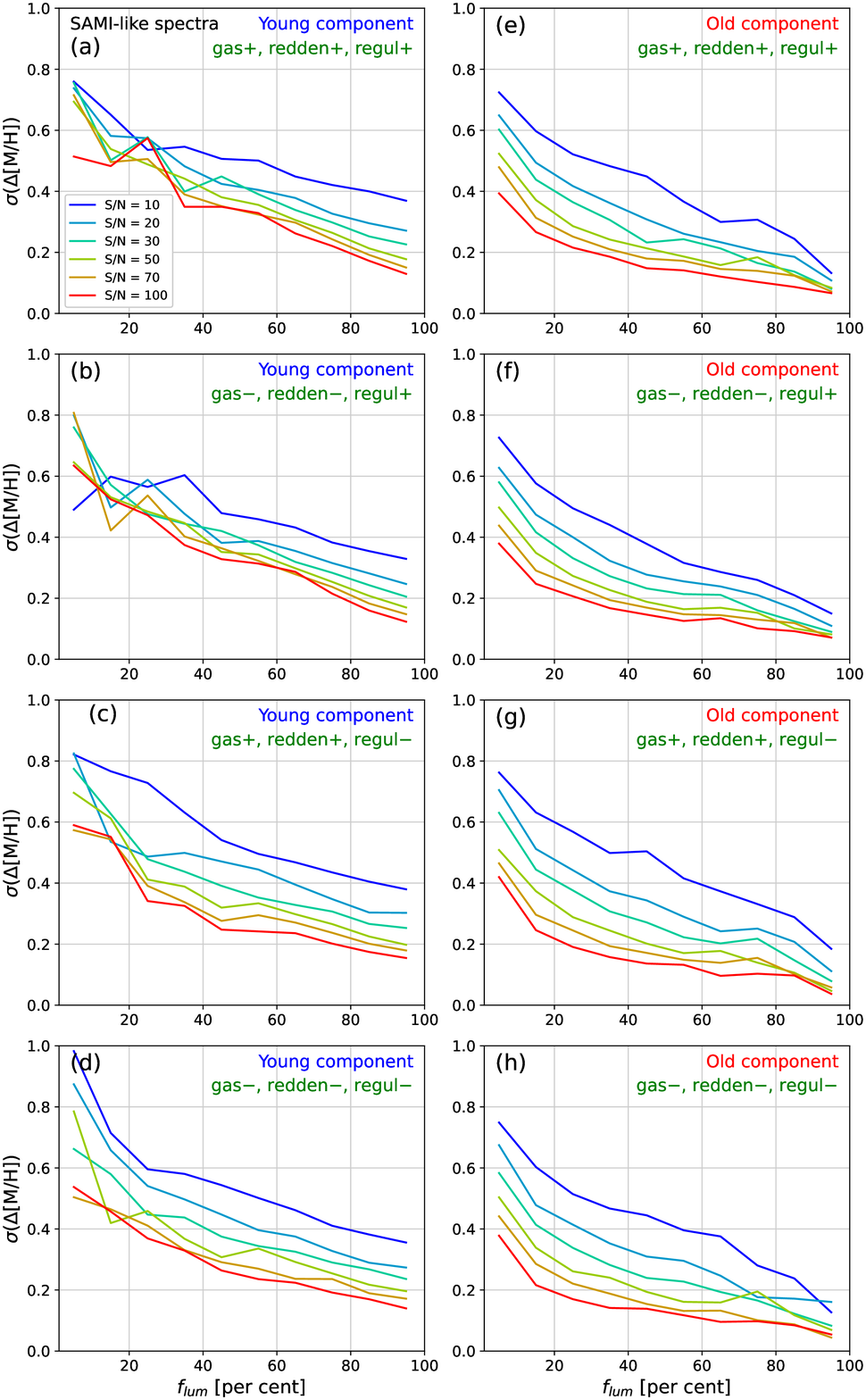}
\caption{The same as Figure~\ref{met1}, but when the pPXF fitting is done for the SAMI-like spectra. \label{met2}}
\end{figure*}

\begin{figure*}
\centering
\includegraphics[width=1.6\columnwidth]{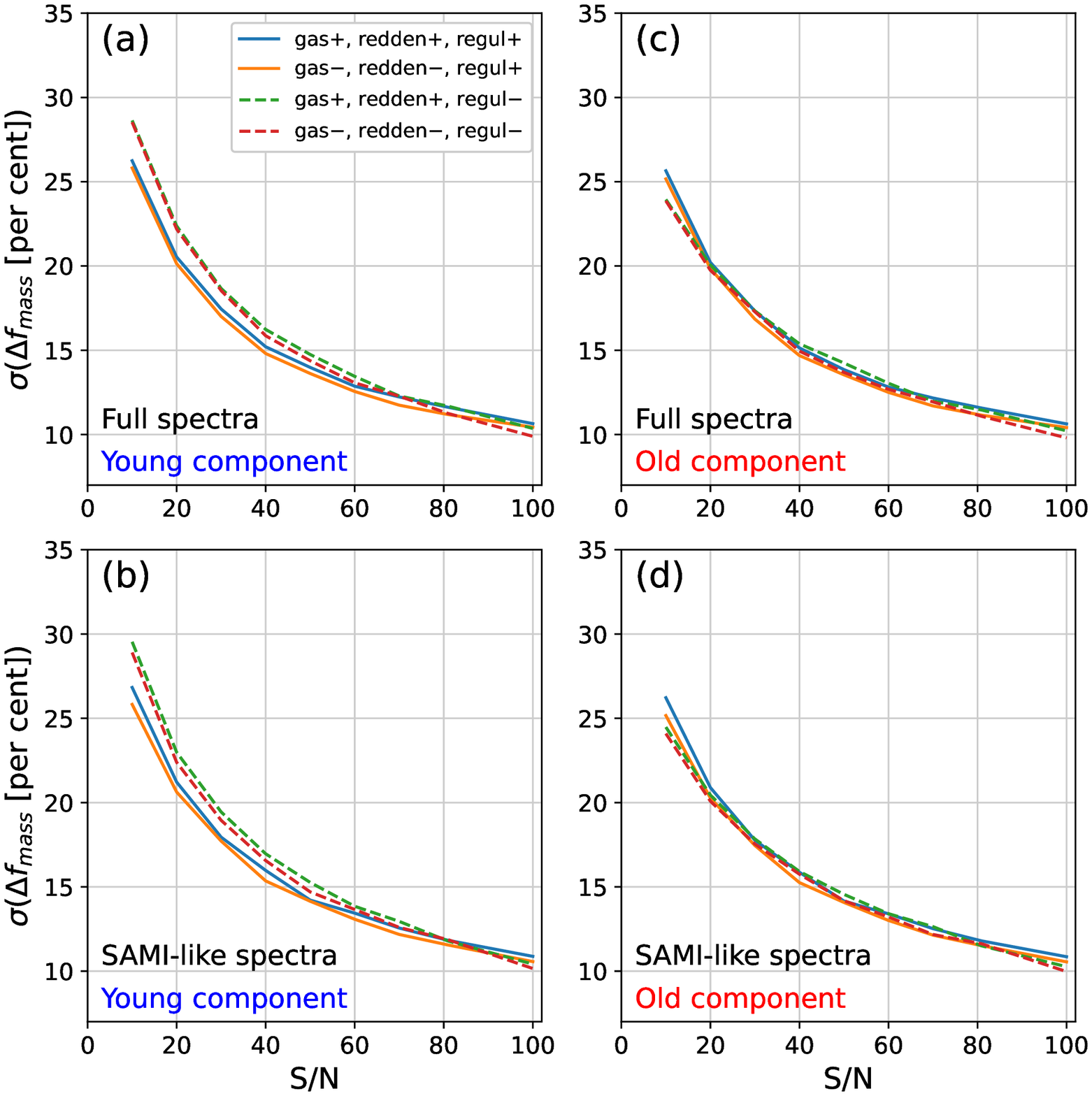}
\caption{Standard deviation of $\Delta f_{\textrm{mass}}$ as a function of S/N, with various options of gas emission, dust reddening, and regularisation: (a) for young components with the full spectra; (b) for young components with the SAMI-like spectra; (c) for old components with the full spectra; and (d) for old components with the SAMI-like spectra. \label{fra2}}
\end{figure*}

\subsection{Comparison Between Input and Output: Age-Divided Mean Stellar Populations (ADPs)}\label{compm}

It is not easy to quantitatively determine how similar a star formation and chemical evolution history is to another one, because such a history is basically a two-dimensional array containing the distribution of mass fraction on the age versus metallicity plane. 
Figure~\ref{inout} presents an example of how the template composition is reproduced by the pPXF fitting.
Since the regularisation option is on in this fitting, the output templates tend to be largely extended unlike the input templates. Nevertheless, one may feel that the overall distribution of the output templates resembles that of the input ones. However, such impression is not easy to be expressed as quantitative indicators showing `how similar the input and output templates are'.

Instead of trying to perfectly compare the two-dimensional arrays, we devised a simple but practical approach: comparison of age-divided mean stellar populations (ADPs). In this approach, we divide the stellar populations of a galaxy into two components: young and old components, with a criterion of log(age/yr) = 9.5 (age $\approx 3.2$ Gyr). This criterion is our choice in this paper, but it may be differently chosen according to the purpose of research. For example, if more tight definition of young components is necessary, one may choose a criterion of log(age/yr) = 9.0, but it needs to be kept in mind that the smaller fraction of young components results in the lower reliability of their estimated mean stellar populations (see Section~\ref{snflum}).
For each component, we estimate the mean stellar population (age and metallicity) and its mass fraction, which will be compared between the input and output values. The age and metallicity distributions of the ADPs of our 10,000 random spectra (input) are presented in Figure~\ref{input}(d) and (e).

This method may be too simple to compare every detail between input and output, and thus may not be suitable for some specific purposes. 
Note that we do not focus on the perfect recovery of the full history of star formation and chemical evolution, when using ADPs. In the full spectrum fitting, some input components may be underestimated or even missing, while some input components may be overestimated or even ghost components may appear, as exemplarily compared in Figure~\ref{inout}. Such inaccuracy is simply dealt with as the statistical uncertainty of ADPs, in this paper.
Nevertheless, ADPs constrain the star formation and chemical evolution histories of galaxies much better than the comparison of the total mean stellar populations without age division. How this approach can improve our understanding of a galaxy formation history will be discussed in section~\ref{app} with a few examples.

\begin{figure*}
\centering
\includegraphics[width=1.66\columnwidth]{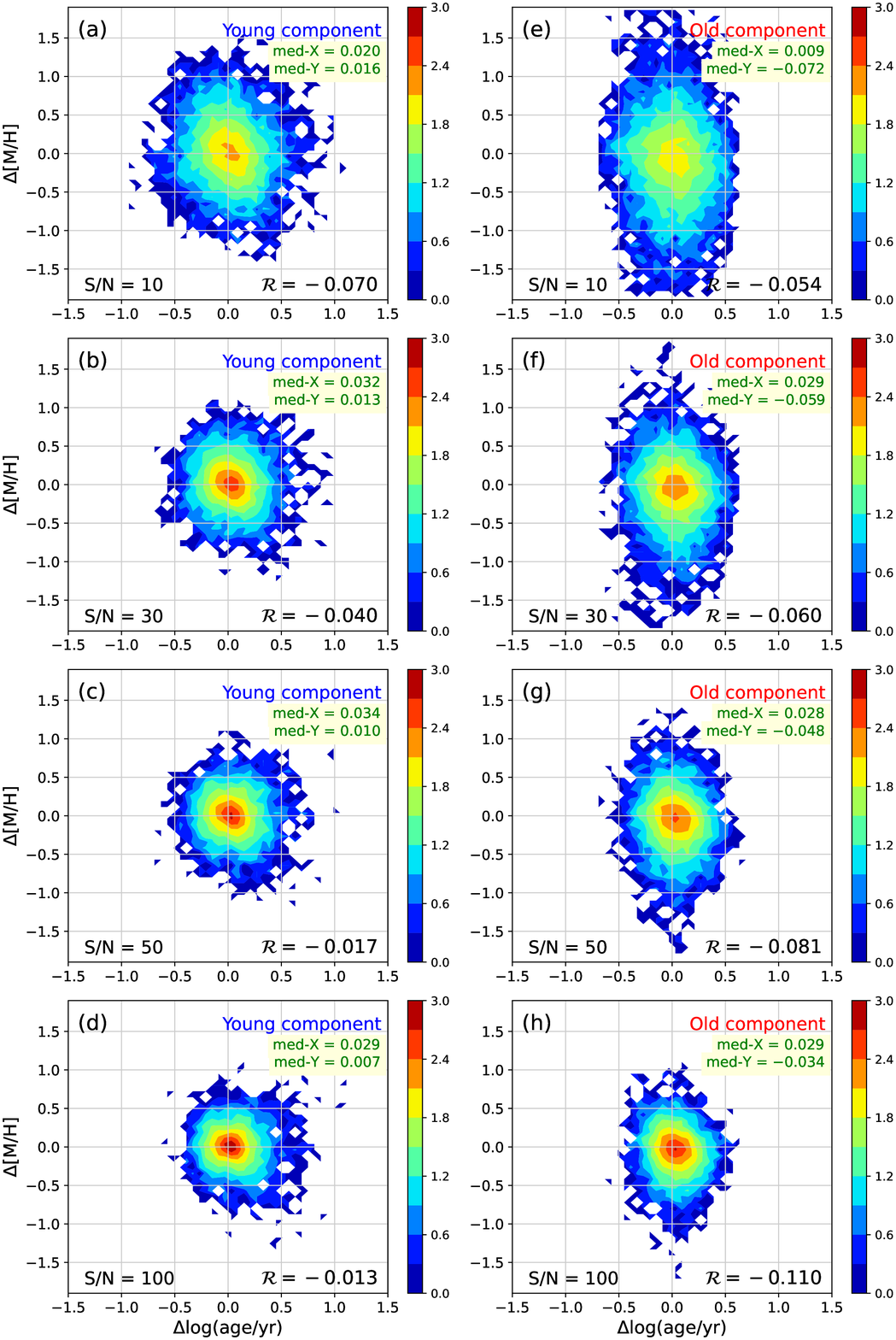}
\caption{Correlation between $\Delta$[M/H] and $\Delta$log(age/yr) of the young and old stellar components according to S/N, presented by contour maps of log(the number of data points). Pearson correlation coefficient ($\mathcal{R}$) is denoted in each panel. Most p-values are extremely small ($\ll 0.0001$), while the p-values for (c) and (d) are 0.0837 and 0.1925, respectively. The median values of X- and Y-axis parameters (med-X and med-Y) are also denoted in each panel. \label{errdep1}}
\end{figure*}

\begin{figure*}
\centering
\includegraphics[width=1.66\columnwidth]{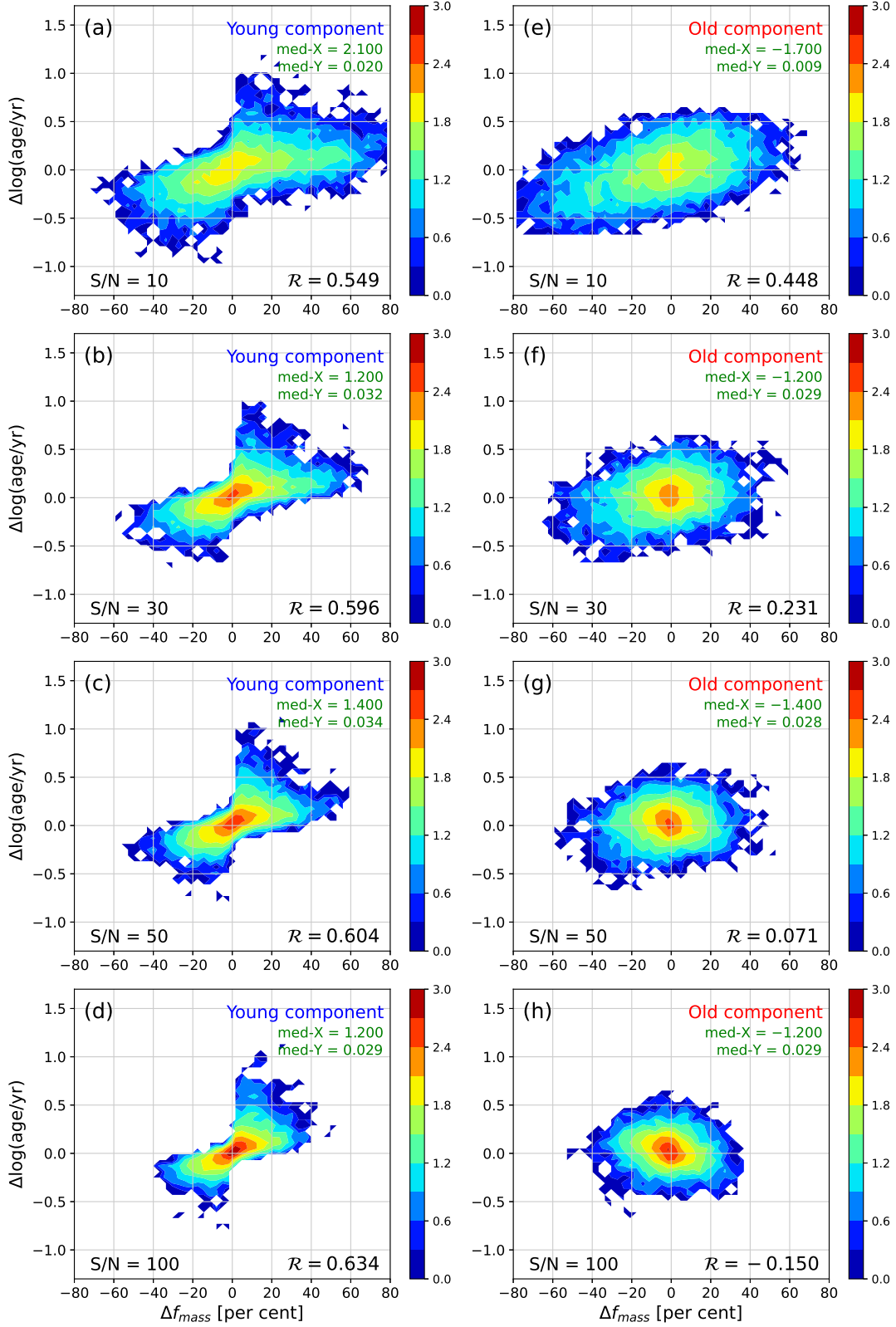}
\caption{Correlation between $\Delta$log(age/yr) and $\Delta f_{\textrm{mass}}$ of the young and old stellar components according to S/N, presented by contour maps of log(the number of data points). Pearson correlation coefficient ($\mathcal{R}$) is denoted in each panel. All p-values are extremely small ($\ll 0.0001$). The median values of X- and Y-axis parameters (med-X and med-Y) are also denoted in each panel. \label{errdep2}}
\end{figure*}

\begin{figure*}
\centering
\includegraphics[width=1.66\columnwidth]{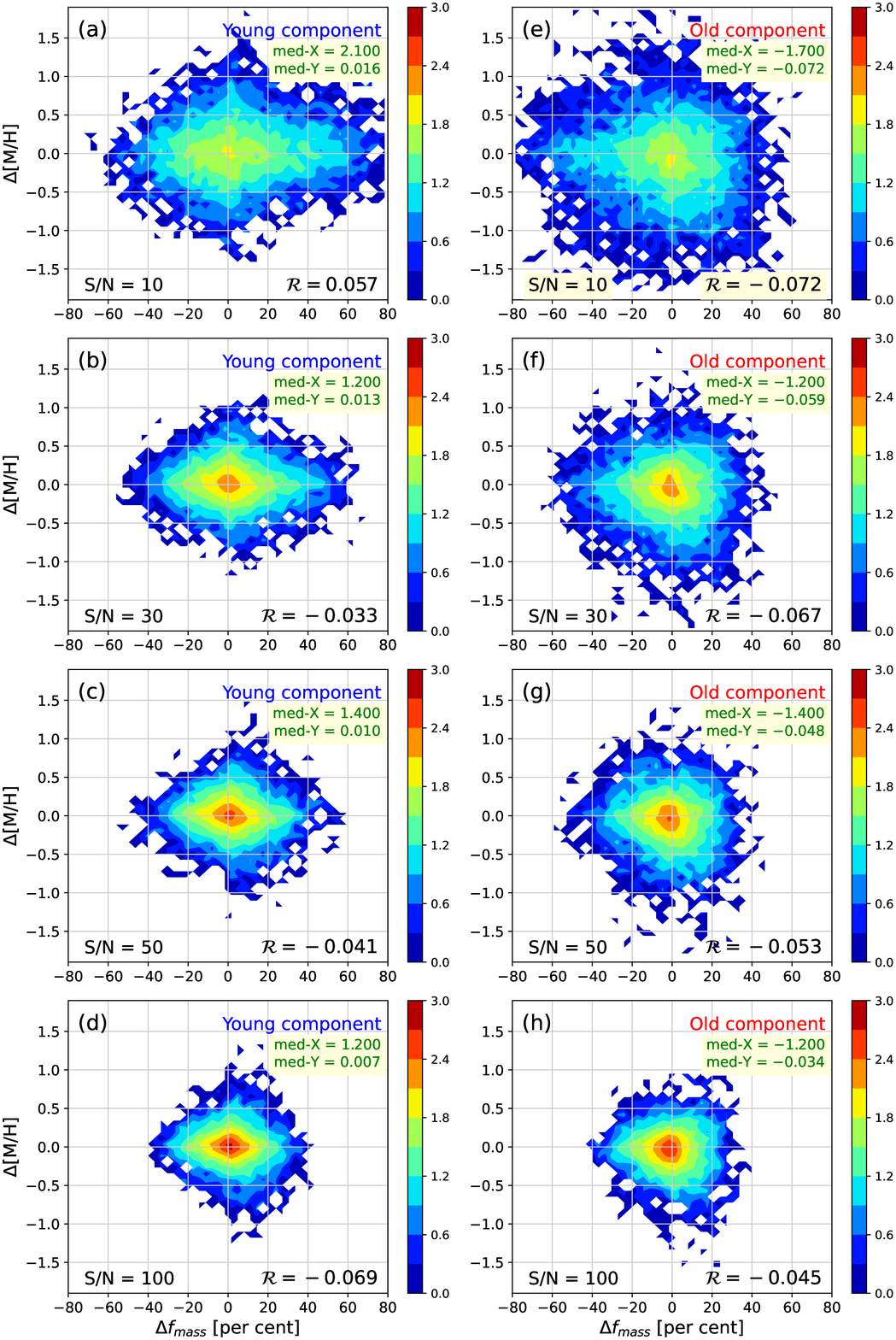}
\caption{Correlation between $\Delta$[M/H] and $\Delta f_{\textrm{mass}}$ of the young and old stellar components according to S/N, presented by contour maps of log(the number of data points). Pearson correlation coefficient ($\mathcal{R}$) is denoted in each panel. Most p-values are extremely small ($\ll 0.0001$), while the p-value for (b) is 0.0011. The median values of X- and Y-axis parameters (med-X and med-Y) are also denoted in each panel. \label{errdep3}}
\end{figure*}

\begin{figure*}
\centering
\includegraphics[width=1.66\columnwidth]{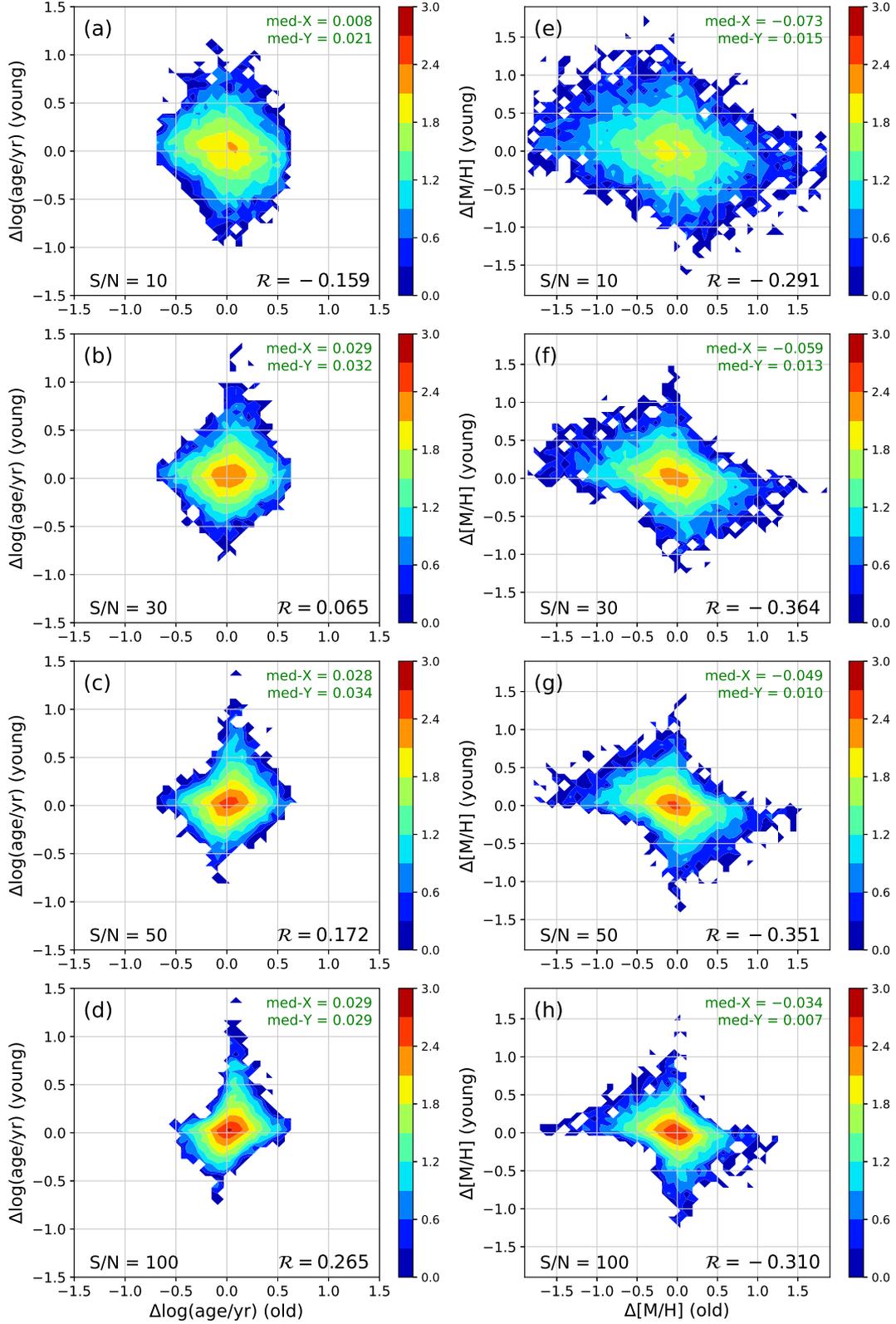}
\caption{(a) - (d) Correlation between $\Delta$log(age/yr) of the young stellar components and $\Delta$log(age/yr) of the old stellar components according to S/N. (e) - (f) Correlation between $\Delta$[M/H] of the young stellar components and $\Delta$[M/H] of the old stellar components according to S/N, presented by contour maps of log(the number of data points). Pearson correlation coefficient ($\mathcal{R}$) is denoted in each panel. All p-values are extremely small ($\ll 0.0001$). The median values of X- and Y-axis parameters (med-X and med-Y) are also denoted in each panel. \label{errdep4}}
\end{figure*}

\begin{figure}
\centering
\includegraphics[width=0.9\columnwidth]{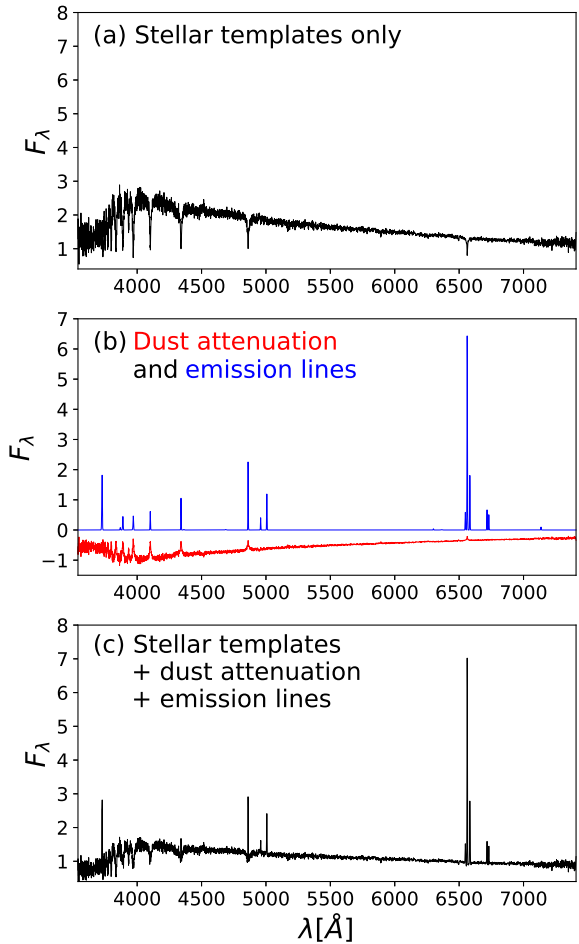}
\caption{An example of applying dust attenuation and emission lines to an artificial spectrum. (a) An artificial spectrum consisting of five simple stellar populations, with S/N = 30. (b) Dust attenuation to be subtracted from the artificial spectrum \citep[red line; E($B-V$) = 0.1 with the extinction law of][]{cal00}, and emission lines to be added (blue line; the ratio of H$\alpha$ amplitude to continuum = 5, using the emission line template of SAMI-70114. (c) The finally combined spectrum. \label{emexsp}}
\end{figure}

\begin{figure*}
\centering
\includegraphics[width=1.9\columnwidth]{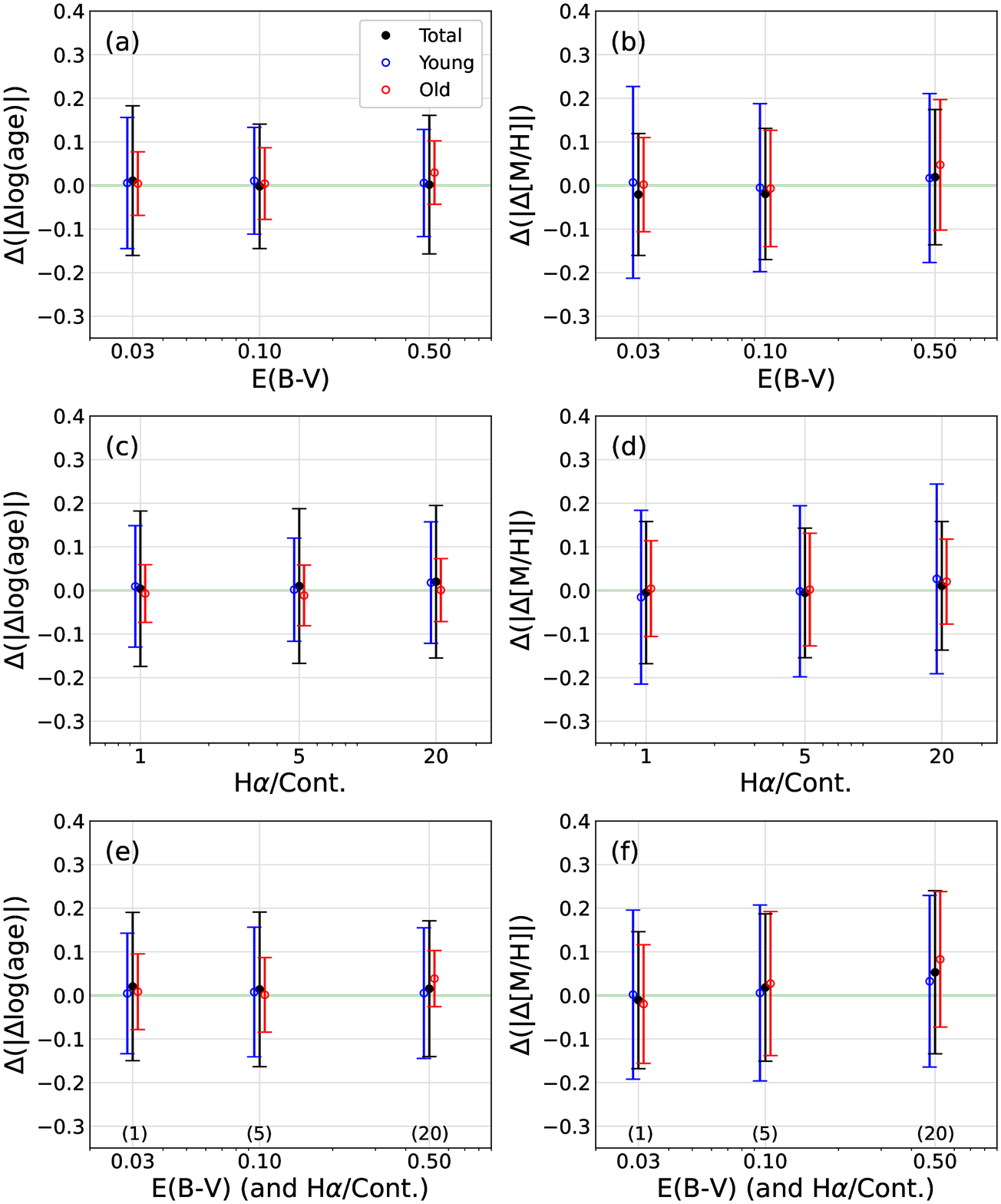}
\caption{The influence of dust attenuation and emission lines to $\Delta$log(age) and $\Delta$[M/H]. Where $\Delta$(|$\Delta X$|) is the difference of the absolute value of $\Delta X$ (`with dust attenuation or emissions' $-$ `without dust attenuation or emissions'; $X$ = log(age) or [M/H]), the filled or open circles show the mean $\Delta$(|$\Delta X$|), while the errorbars show the standard deviation. (a) $\Delta$(|$\Delta$log(age)|) when E($B-V$) = 0.03, 0.1, and 0.5, without emission lines, for total mean stellar population (black), young-component mean stellar population (blue), and old-component mean stellar population (red). (b) $\Delta$(|$\Delta$[M/H]|) when E($B-V$) = 0.03, 0.1, and 0.5, without emission lines. (c) - (d) The same as (a) and (b), but when the ratio of H$\alpha$ amplitude to continuum (H$\alpha$ / Cont.) = 1, 5, and 20, without dust attenuation. (e) - (f) The same as (a) and (b), but when dust attenuation and emission lines are applied together: E($B-V$) = 0.03 \& H$\alpha$ / Cont. = 1; E($B-V$) = 0.1 \& H$\alpha$ / Cont. = 5; and E($B-V$) = 0.5 \& H$\alpha$ / Cont. = 20. For the mean population of young or old components, the spectra with $f_{\textrm{lum}}$ of each component $<30\%$ were not used. \label{emexplt}}
\end{figure*}

\section{RESULTS}\label{result}

In this section, we compare the total mean stellar populations as well as ADPs between the input and output spectra. The degeneracy between age, metallicity and mass fraction is also examined.

\subsection{Total Mean Stellar Populations}\label{totpop}

Before comparing ADPs between input and output spectra, we first analyse how well the total mean stellar populations (not divided by age) are reproduced by full spectrum fitting.
Figure~\ref{singlesn} presents the dispersion (standard deviation) of $\Delta$ (= output $-$ input) as a function of S/N, for various settings of the pPXF runs. Overall, the dispersion of $\Delta$ significantly decreases as S/N increases, while the differences between using different pPXF settings are not large; for example, the difference of $\sigma(\Delta$[M/H]) between the `gas+, redden+, regul+' and `gas$-$, redden$-$, regul$-$' options is only 0.014 at S/N=30.

The gas emission and dust reddening options do not seem to significantly influence the reliability in age estimation, whereas the dispersion in $\Delta$[M/H] estimation appears to be slightly reduced when those options are off. However, since we did not include gas emission or dust attenuation in our input spectra, these results may be different for real galaxies with gas and dust. Turning on the regularisation option seems to reduce $\Delta$ when S/N is relatively low ($\lesssim 50$), while it does not when S/N is high. 
This may be related to the fact that the templates in the input spectra are discrete rather than extended (five simple stellar populations), which may make the output more accurate when S/N is high with regularisation off, but may work oppositely when S/N is low. 
To confirm this speculation, tests for more complicated input spectra (with each population more realistically dispersed) may be necessary, but we do not conduct such tests in this paper, because the difference is negligible compared to the statistical uncertainties (mostly smaller than 0.01 dex). The difference between the full spectra and the SAMI-like spectra appears to be even smaller.

In addition, we examine whether $\Delta$log(age/yr) and $\Delta$[M/H] are correlated to each other, because it is well known that age and metallicity tend to be closely entangled in the estimation of stellar populations \citep[so-called age-metallicity degeneracy; e.g.,][]{sil94,fer00,kav07}. For example, a young and metal-rich population can have similar broad spectral properties to an old and metal-poor population, particularly if the data quality is not good enough. However, it is known that such degeneracy can be largely mitigated  when the observed spectrum has high S/N \citep{san11,con13}.

To quantitatively check such a degeneracy effect, Figure~\ref{agemet} presents the correlations between $\Delta$log(age/yr) and $\Delta$[M/H] in the total mean stellar populations, according to S/N. The two $\Delta$ values appear to be very weakly anti-correlated, which shows a small hint that age may be slightly underestimated if metallicity is overestimated and vice versa. However, since the strength of such an anti-correlation is very weak (Pearson correlation coefficient $|\mathcal{R}|<0.2$ when S/N $\geq 30$), the age-metallicity degeneracy does not have a strong effect on the estimation of the total mean stellar populations using the full spectrum fitting. Thus, if we secure sufficient S/N (e.g., $\geq 30$), we may safely assume that S/N is the solely critical factor determining the reliability of total mean stellar populations.

\subsection{Age-Divided Mean Stellar Populations: Dependence on S/N and $f_{\textrm{lum}}$}\label{snflum}

In the comparison of ADPs, age, metallicity, and mass fraction are the three key quantities to constrain the star formation and chemical evolution history. On the other hand, the luminosity fraction ($f_{\textrm{lum}}$) of each component is technically important, because it can directly affect the effective S/N of each component and thus affect the output reliability. For example, even if two galaxies A and B have the same mass fraction of young stars, their luminosity fractions may not be the same, because the mass-to-light ratio strongly depends on age, as shown in Figure~\ref{masslum}. If the young component of galaxy A is 0.1 Gyr old, while that of galaxy B is 1 Gyr old, $f_{\textrm{lum}}$ of the former will be almost ten times larger than that of the latter. Then, even though the two galaxy spectra have the same total S/N, the young component of galaxy A has much higher effective S/N than the young component of galaxy B, which may result in higher reliability of the outputs for the former.
Thus, luminosity fraction is technically influential, while mass fraction is important in its physical implication.

Figure~\ref{age1} presents how S/N and $f_{\textrm{lum}}$ affect the dispersion of $\Delta$log(age/yr), respectively for young and old stellar components. Here and after, we use the \emph{output} luminosity fraction as $f_{\textrm{lum}}$, in consideration of the practical convenience when applying our results to actual investigation of observational data.
Basically, the dispersion of $\Delta$ tends to decrease as S/N and $f_{\textrm{lum}}$ increase. However, when $f_{\textrm{lum}}$ is small (e.g., $\lesssim 30$ per cent), the dispersion of $\Delta$ appears to fluctuate for young components (i.e., lower S/N returns smaller dispersion sometimes), which is because the number of data points is relatively small there. Overall, the dispersions of $\Delta$log(age/yr) for young components are clearly larger than those for old components, when S/N and $f_{\textrm{lum}}$ are fixed; for example, $\sigma$($\Delta$log(age/yr)) $\sim0.25$ versus 0.13 for young and old components respectively, at S/N = 30, $f_{\textrm{lum}}$ = 50 per cent, and all pPXF options on. This may be influenced by the fact that the noise function rapidly increases as wavelength decreases at $\lambda < 4000$ {\AA}, which is an important range to precisely determine the ages of young stellar populations. It is noted that the impact of $f_{\textrm{lum}}$ seems to be even larger than that of S/N. For example, the difference in the dispersion of $\Delta$log(age/yr) of young components between S/N = 30 and S/N = 100 appear to be not large ($\lesssim 0.05$), whereas the difference between $f_{\textrm{lum}} = 50$ and 100 per cent is larger than 0.1.

The difference between the on/off of gas emission and dust reddening options in pPXF seems to be minor at least in our sample spectra that do not contain emission lines or internal dust contents. However, the regularisation option seems to result in more visible (but still small) differences between on (four upper panels in Figure~\ref{age1}) and off (four lower panels in Figure~\ref{age1}). It is not easy to assert which choice returns better results between the regularisation on and off, because the effect by this option does not appear to be consistent over S/N. Overall, when the regularisation is off, the dependence on S/N seems to be larger than the case of regularisation on. In other words, the regularisation may reduce the dispersion of $\Delta$ when S/N is low, but it does not necessarily do when S/N is high (e.g., Figure~\ref{age1}(a) and Figure~\ref{age1}(c)).
We also tested the same plots as Figure~\ref{age1} for SAMI-like spectra in Figure~\ref{age2}. As a result, the difference between the full spectra and the SAMI-like spectra appears to be almost negligible.

Figures~\ref{met1} and \ref{met2} show the results of the same test for the dispersion of $\Delta$[M/H]. The overall trends are similar to the cases of the dispersion of $\Delta$log(age/yr), but the dispersion of $\Delta$[M/H] appears to depend on S/N more strongly than the dispersion of $\Delta$log(age/yr), especially for old stellar components. As $\Delta$log(age/yr) does, $\Delta$[M/H] tends to show larger dispersion for young components than for old components, too; for example, $\sigma(\Delta$[M/H]$) \sim0.40$ and 0.23, respectively, at S/N = 30, $f_{\textrm{lum}}=50$ per cent, and all pPXF options on.
This may be partially because young populations have weaker metal absorption lines than old populations in the wavelength range considered in this paper, which makes it more difficult to accurately measure the metallicity of young populations than that of old populations when the S/N and $f_{\textrm{lum}}$ are fixed.

Figure~\ref{fra2} presents how $\Delta f_{\textrm{mass}}$ depends on S/N. The dependence on $f_{\textrm{lum}}$ is not considered here, because $f_{\textrm{mass}}$ and $f_{\textrm{lum}}$ are closely entangled quantities. Compared to $\Delta f_{\textrm{mass}}$ of young components, $\Delta f_{\textrm{mass}}$ of old components appears to be less influenced by the regularisation option. The differences between the full spectra and the SAMI-like spectra is negligible.

\subsection{Age-Divided Mean Stellar Populations: Degeneracy}\label{degen}

As done for the total mean stellar populations in Section~\ref{totpop}, we check the effects of the age-metallicity degeneracy in the estimation of ADPs, too. In Figure~\ref{errdep1}, $\Delta$[M/H] and $\Delta$log(age/yr) show no meaningful correlations, when divided into the young and old stellar components, regardless of S/N. In the case of ADPs, however, an additional factor needs to be considered: $\Delta f_{\textrm{mass}}$. Figures~\ref{errdep2} and \ref{errdep3} present the $\Delta$log(age/yr) versus $\Delta f_{\textrm{mass}}$, and the $\Delta$[M/H] versus $\Delta f_{\textrm{mass}}$ correlations, respectively. In Figure~\ref{errdep2}, the $\Delta$log(age/yr) of the young stellar components clearly depends on their $\Delta f_{\textrm{mass}}$ regardless of S/N, in the context that the mean age of the young stellar components tends to be overestimated when their mass fraction is overestimated ($\mathcal{R} \sim 0.6$). On the other hand, for the old stellar components, such a correlation appears only when S/N is very low (S/N = 10). Meanwhile, in Figure~\ref{errdep3}, $\Delta$[M/H] does not show any meaningful correlation with $\Delta f_{\textrm{mass}}$, for either of young or old stellar components.

In addition to the degeneracy among the three parameters, we also check if there is any degeneracy between young and old stellar components.
Figure~\ref{errdep4} presents the correlations between the young and old stellar components, in $\Delta$log(age/yr) and in $\Delta$[M/H]. While the correlations are not so significant ($|\mathcal{R}| < 0.27$) in $\Delta$log(age/yr), the $\Delta$[M/H] values of the young and old components appear to be marginally anti-correlated ($-0.36 \lesssim \mathcal{R} \lesssim -0.29$).
Note that this comparison is not necessary in $\Delta f_{\textrm{mass}}$, because it is obvious that they are anti-correlated by definition: $\Delta f_{\textrm{mass}}$ (young) $= -\Delta f_{\textrm{mass}}$ (old).

\subsection{Dust Attenuation and Emission Lines}\label{emex}

In this section, we briefly examine how much dust attenuation and emission lines can influence the ADP estimation. However, we do not produce complete tables for it, because the number of possible combinations of those effects is too large and a full examination is beyond the scope of this paper. Here, we test them under some specific assumptions: a specific attenuation law and a specific emission line template, in limited conditions.

We tested three cases of dust attenuation, assuming the extinction law of \citet{cal00} and parametrised by the colour excess: E($B-V$) = 0.03, 0.1, and 0.5.
For emission lines, we extracted the emission lines from SAMI-70114, a dwarf galaxy that has active star formation at its center.
Note that the attenuation curves and emission line templates are not universal, and thus the test results from this scheme cannot be safely applied to all kinds of galaxies. Figure~\ref{emexsp} shows an example of applying dust attenuation and emission lines to an artificial galaxy spectrum.

To quantify the impact of dust attenuation and emission lines, we define a new parameter, as follows:
\begin{equation}
\begin{array}{rr}
\Delta (|\Delta X|) = |\Delta X| \:\textrm{with dust attenuation and/or emission lines} \\
- |\Delta X| \:\textrm{without dust attenuation or emission lines},
\end{array}
\end{equation}
where $X$ is log(age) or [M/H]. We estimated $\Delta (|\Delta X|)$ for individual spectrum, in the following cases:
\begin{itemize}
\item E($B-V$) = 0.03, 0.1 and 0.5, without emission lines
\item H$\alpha$ / Cont. = 1, 5, and 20, without dust attenuation
\item E($B-V$) = 0.03 \& H$\alpha$ / Cont. = 1; E($B-V$) = 0.1 \& H$\alpha$ / Cont. = 5; and E($B-V$) = 0.5 \& H$\alpha$ / Cont. = 20;
\end{itemize}
where H$\alpha$ / Cont. is the ratio of H$\alpha$ amplitude to continuum.
After running pPXF (with gas+, reddening+, and regularisation+) for these nine cases, we calculated the mean value and standard deviation of $\Delta (|\Delta$log(age)|) and $\Delta (|\Delta$[M/H]|) in each case. For the statistics of young (old) components, only the spectra with $f_{\textrm{lum}}$ of young (old) components $\ge30\%$ were used.

The results are presented in Figure~\ref{emexplt}.
Overall, the mean $\Delta (|\Delta X|)$ tends to be very small (mostly smaller than 0.02 dex, except for a couple of extreme cases), while the standard deviations are much larger than the mean values. The amplitude of the standard deviation seems to be approximately consistent with $\sigma(\Delta X)$ when S/N = 30 in Figures~\ref{age1}-\ref{met2}. These results indicate that the influence of dust attenuation and emission lines is mostly insignificant, compared to the basic uncertainty of ADP estimation. In an extreme case (E($B-V$) = 0.5 and H$\alpha$ / Cont. = 20), however, $\Delta (|\Delta$[M/H]|) is as large as $0.083\pm0.155$ (old components), which shows that exceptionally large dust attenuation and emission lines may visibly increase ADP uncertainty, particularly the old-component metallicity. Nevertheless, in most moderate cases, we do not need to consider the impact of dust attenuation and emission lines too much; for example, when E($B-V$) = 0.1 and H$\alpha$ / Cont. = 5, $\Delta (|\Delta$log(age)|) = $0.014\pm0.177$, $0.008\pm0.149$, and $0.001\pm0.086$; and $\Delta (|\Delta$[M/H]|) = $0.018\pm0.169$, $0.006\pm0.202$, and $0.027\pm0.165$, for the total, young-component, and old-component mean stellar populations, respectively.

\section{DISCUSSION}\label{discuss}

\subsection{Reliability and Degeneracy}\label{keys}

In the estimation of total mean stellar populations, S/N appears to significantly affect the reliability. 
In Figure~\ref{singlesn}, the dispersions of $\Delta$log(age/yr) and $\Delta$[M/H] strongly depend on S/N, with relatively small differences between various pPXF options. Based on these results, the empirical reliability of the total mean stellar populations of observed galaxies estimated from full spectrum fitting can be easily interpolated.
The degeneracy between age and metallicity does not seem to be serious unless S/N is too low, as Figure~\ref{agemet} shows the $\Delta$[M/H] versus $\Delta$log(age/yr) plots almost roundly scattered when S/N $\gtrsim 30$.

For ADPs, however, the situation is more complicated, because one more parameter needs to be considered other than age and metallicity: the mass fractions of young and old stellar components. In this case, the degeneracy may not be only between age and metallicity, but may include mass fraction as third parameter. Indeed, in Figure~\ref{errdep2}, $\Delta$log(age/yr) appears to be obviously correlated with $\Delta f_{\textrm{mass}}$ for young stellar components.
This may be a natural result, in the regard that very young stars with a small mass fraction are difficult to distinguish from moderately young stars with a moderate mass fraction, when they are mixed with old stars with a large mass fraction. It is noted that this degeneracy is not mitigated at all even at the highest S/N ($\sim 100$), which shows how difficult the exact reproduction of a star formation history from full spectrum fitting is.
On the other hand, such a strong degeneracy is hardly found in other cases. Particularly, unlike $\Delta$log(age/yr), it appears that $\Delta$[M/H] shows almost no correlation with $\Delta f_{\textrm{mass}}$. This may be also a natural result, because our method divides stellar components with an age cut, not with a metallicity cut, and thus, the mass fractions according to divided ages may be orthogonal (or at least less influential) to metallicity distribution.

All data tables for Figures~\ref{singlesn}, \ref{age1} - \ref{met2} and \ref{fra2} are available as online material, which can be directly applied to the analysis of the SAMI Galaxy Survey data, with caveats that will be discussed in section~\ref{caveat}. In the case of another spectroscopic data, the instrumental features of which may be significantly different from SAMI, the tables for the dependency of $\Delta$ dispersion on S/N and $f_{\textrm{lum}}$ need to be estimated with a new set of random spectra with the same specifications as the actually observed data. In the next two subsections, we discuss several points to consider in applying our results, and present the exemplary analysis of three SAMI galaxies with ADPs.

\subsection{Caveats}\label{caveat}

Although our approach is convenient to estimate the empirical uncertainty of the parameters from full spectrum fitting in a simple manner, it needs to be kept in mind that there are more complicated factors affecting $\Delta$ in reality.
First of all, our results are based on the assumption that the stellar library and simple stellar population templates from the MILES models reproduce real spectra sufficiently well. Moreover, the templates we used are obtained from a specific setting of stellar models (e.g., a single-sloped initial mass function), which may be much simpler than the reality. If one suspects the applicability of such simplified models, our whole results may need to be revised by applying more complicated models.

Heavy dust attenuation may seriously distort stellar populations, because it makes the optical colours redder and thus stellar populations look older or more metal-rich unless it is accurately estimated. In other words, dust attenuation causes another degeneracy we should consider in estimating stellar populations.
Although we showed that dust attenuation and emission lines at moderate levels do not significantly influence the ADP estimation (Section~\ref{emex}), it needs to be kept in mind that the ADP uncertainty may be larger than our tables for some extreme galaxies with severe dust attenuation and very strong emission lines (e.g., ultra-luminous infrared galaxies).
Moreover, our experiment in Section~\ref{emex} has been done under very limited conditions and assumptions, while the properties of dust contents in galaxies as well as their emission line features may be much more various and complicated \citep[e.g.,][]{dra84,men10,zaf11,nat13,seo16,hou17,mck18}.

To a galaxy hosting an active galactic nucleus (AGN), it may be dangerous to simply apply our results. 
Although the influence of a type-2 AGN may be limited, a type-1 AGN may dominate the continuum and thus distort the stellar population estimates. Therefore, if the existence of a type-1 AGN is identified in a spectrum, our empirical uncertainty of the total mean stellar populations or ADPs becomes much less reliable.

The degeneracy effects among the parameters appear to be insignificant for most cases, but the correlation between ages and mass fractions of young stellar components needs to consider, as shown in Figure~\ref{errdep2}. For example, if the full spectrum fitting for an observed spectrum returns the existence of extremely young stars with a very small mass fraction, the possibility needs to be considered that the stellar component may be older than the estimated age, with a larger mass fraction. In addition, the weak anti-correlation in $\Delta$[M/H] between young and old components (Figure~\ref{errdep4}) may be worth keeping in mind, too, although the correlations are not as strong as those in Figure~\ref{errdep2}.

Galaxies often show pretty complex internal kinematics, which may influence the estimation of their star formation histories. In this paper, however, we did not consider that respect. Although such effects may not be serious in many cases, extremely unstable kinematics of galaxies (e.g., mergers) may make the ADP uncertainty larger. In addition, since our whole tests followed the abundance pattern of the Milky Way Galaxy (\emph{baseFe} models in the MILES library), the possible impact of various [$\alpha$/Fe] abundance ratios was not examined.

Finally, this paper does not consider extremely young stellar populations. Since the lower limit of ages in the stellar library we used is 63 Myr, star-forming galaxies with a bulk of newly born stars (e.g., 1 $-$ 20 Myr old) cannot be well analysed with our tables. It is known that estimating the metallicity of such extremely young stars is extremely challenging \citep[e.g,][]{con13}.

\subsection{Comparison to a Previous Study}

Here, we discuss how different our study is from \citet{ge18}, which tested the stellar population recovery performance of the softwares pPXF and STARLIGHT \citep{cid05} with an approach similar to ours, to some extent.
Like us, \citet{ge18} tested how well pPXF reproduces input spectra that are based on the MILES library. However, each input spectrum of theirs was mostly a simple stellar population (SSP) spectrum or at most the combination of two SSPs with fixed luminosity ratios (normalizing fluxes to 1 at $\lambda \sim 5500$ {\AA}), whereas we built our input spectra by combining five SSPs with random (but guided) mass ratios as described in Section~\ref{buildspec}.
Moreover, \citet{ge18} used a very simple recipe for metallicity: their 2-SSP spectra were tested only for solar metallicity. On the other hand, we used fully random metallicity within the MILES library for each of the five SSPs to be combined in an input spectrum.
Thus, our input spectra may be much better in simulating realistic galaxy spectra with more various and complex formation histories.

Our method for adding uncertainty to input spectra is also different from that of \citet{ge18}. For the uncertainty of input spectra, \citet{ge18} tested six types of error spectra, five of which have smooth slopes of SSPs with solar metallicity and different ages, and the remaining one has a constant S/N per wavelength pixel (and thus follows the shape of an input spectrum). After checking the effects of the different error spectra, \citet{ge18} finally adopted the flat error spectrum. However, since uncertainty largely depends on not only wavelength but also flux, a flat error spectrum commonly applied to different input spectra with different spectral shapes may not be sufficiently reasonable nor realistic. That is why we defined a noise function, which reflects the dependence of uncertainty on both of wavelength and flux, as described in Section~\ref{addnoise}.

Most importantly, the practical goal of \citet{ge18} was significantly different from ours. What \citet{ge18} tested was how well the full spectrum fitting codes (pPXF and STARLIGHT) recover the mean stellar population of an input spectrum. Because their focus was not on the star formation history, they did not particularly consider the complex combination of various templates in an input spectrum. That is why they used very simple input spectra of single SSPs or 2-SSP combinations.
On the other hand, the goal of our work is to devise a practical method for describing various star formation and chemical evolution histories (and our solution is ADPs), and to examine the statistical reliability of such a method. That is why more various and complex combinations of stellar templates were necessary in this paper.

\subsection{Application to Observed Data}\label{app}

\begin{figure*}
\centering
\includegraphics[width=2.0\columnwidth]{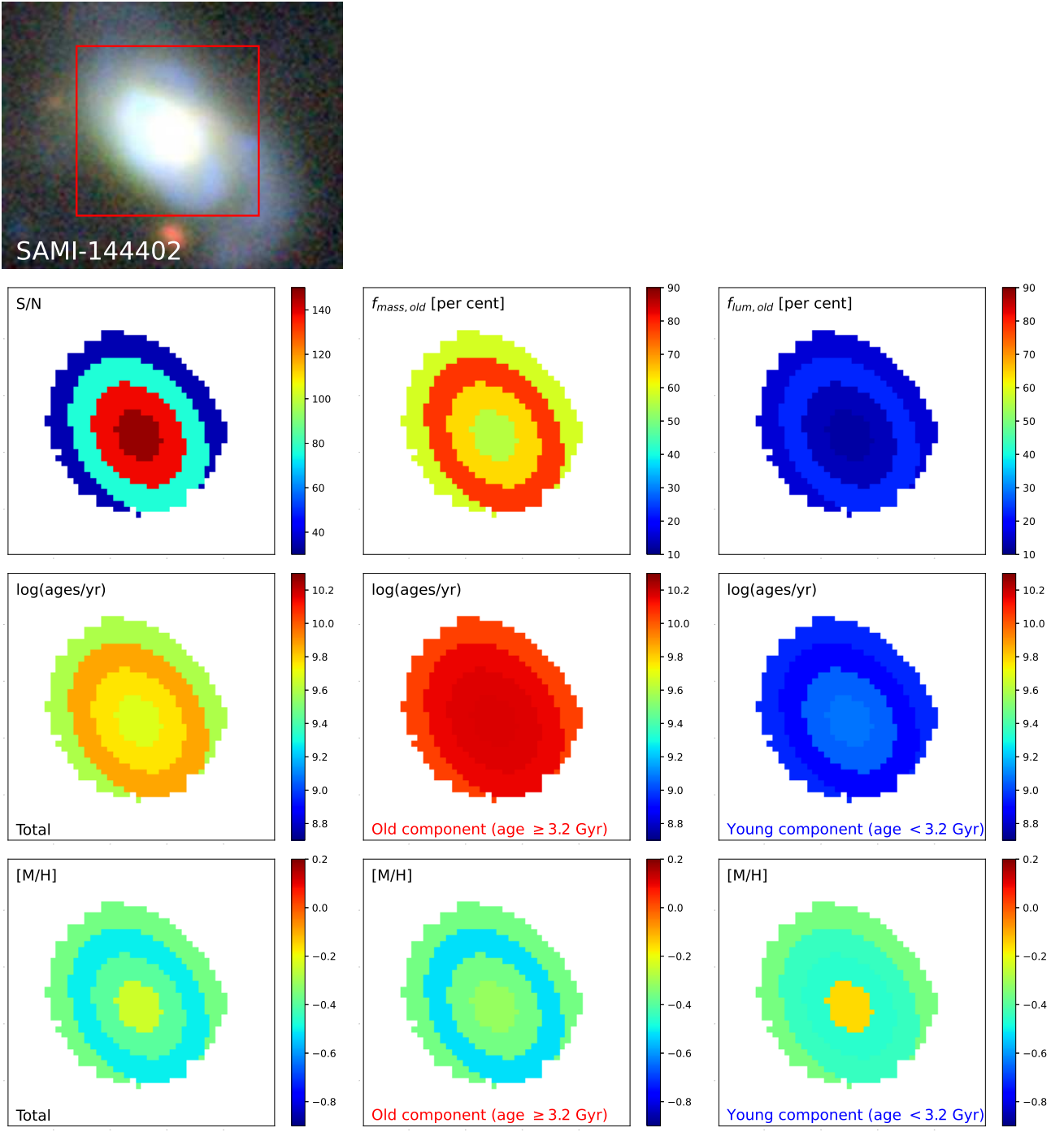}
\caption{An example of the ADP gradient analysis for SAMI-144402. The panels in the first row below the atlas image show the maps of S/N, mass fraction of old components, and luminosity fraction of old components, from left to right. The panels in the second and third rows show the mean log(age/yr) and mean [M/H], respectively: left panels for the total mean stellar populations, middle panels for old components ($\geq 10^{9.5}$ yr $\approx 3.2$ Gyr), and right panels for young components ($< 10^{9.5}$ yr). \label{exam1}}
\end{figure*}

\begin{figure*}
\centering
\includegraphics[width=2.0\columnwidth]{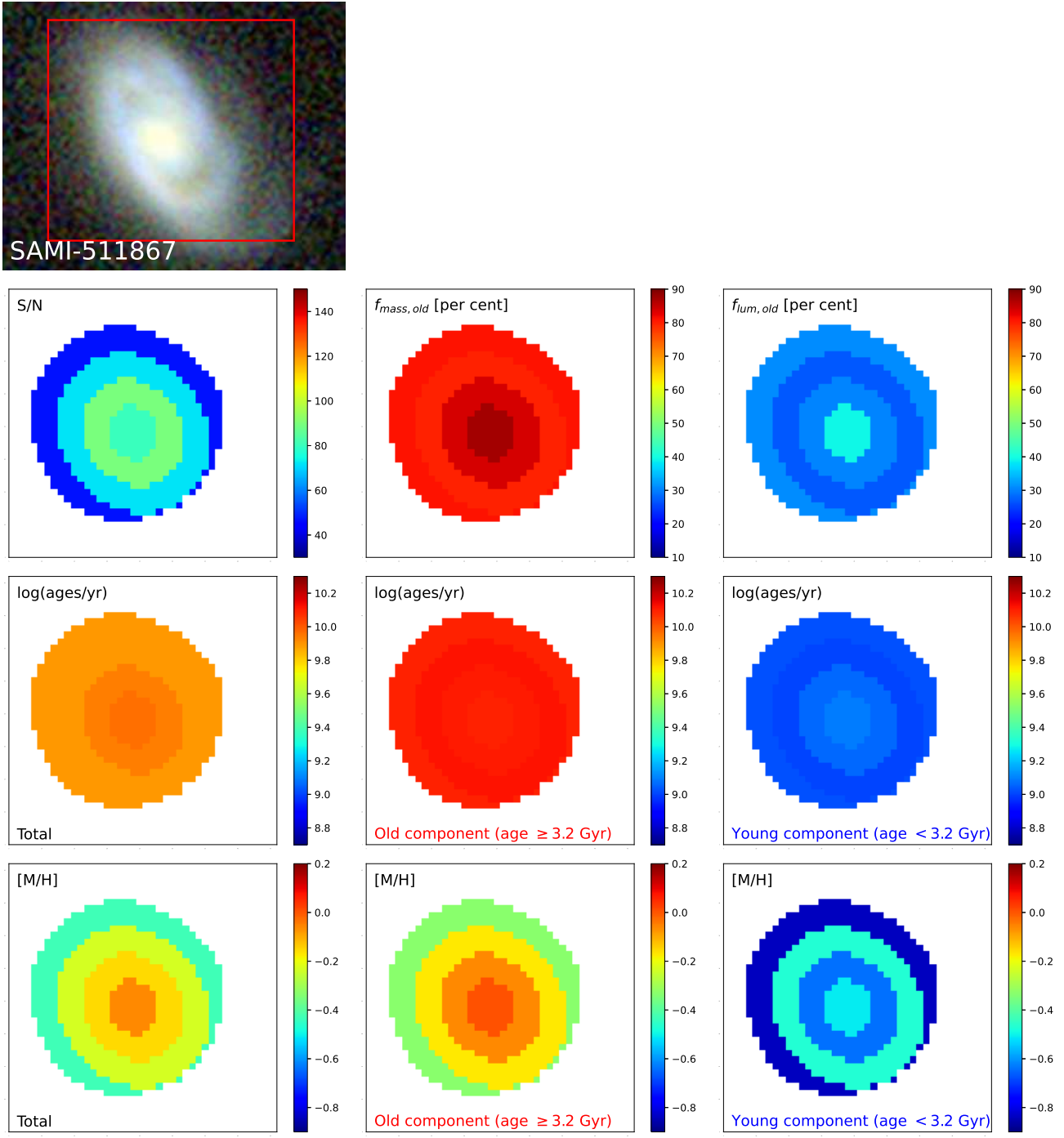}
\caption{An example of the ADP gradient analysis for SAMI-511867.\label{exam2}}
\end{figure*}

\begin{figure*}
\centering
\includegraphics[width=2.0\columnwidth]{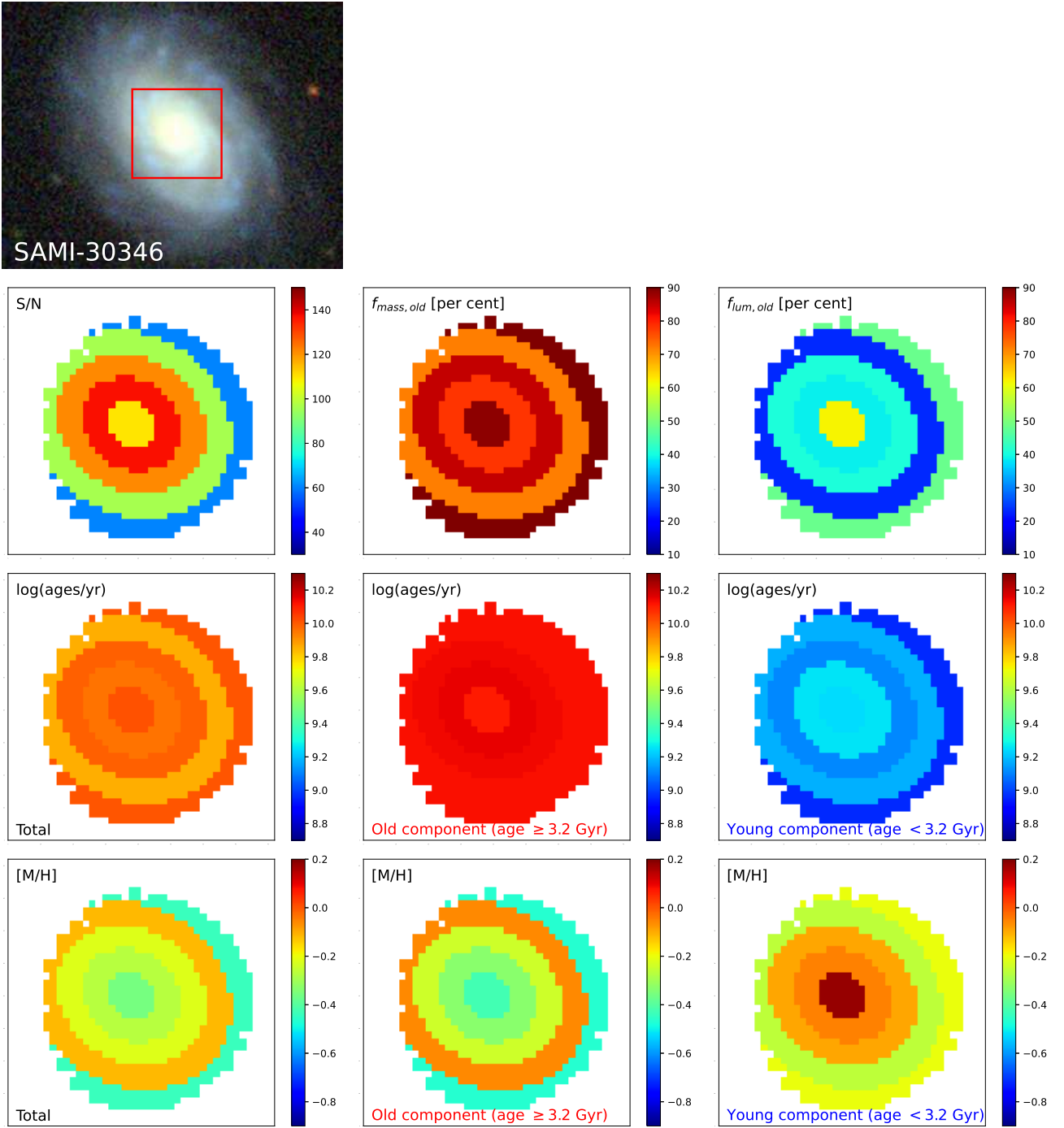}
\caption{An example of the ADP gradient analysis for SAMI-30346. \label{exam3}}
\end{figure*}

\begin{figure*}
\centering
\includegraphics[width=2.0\columnwidth]{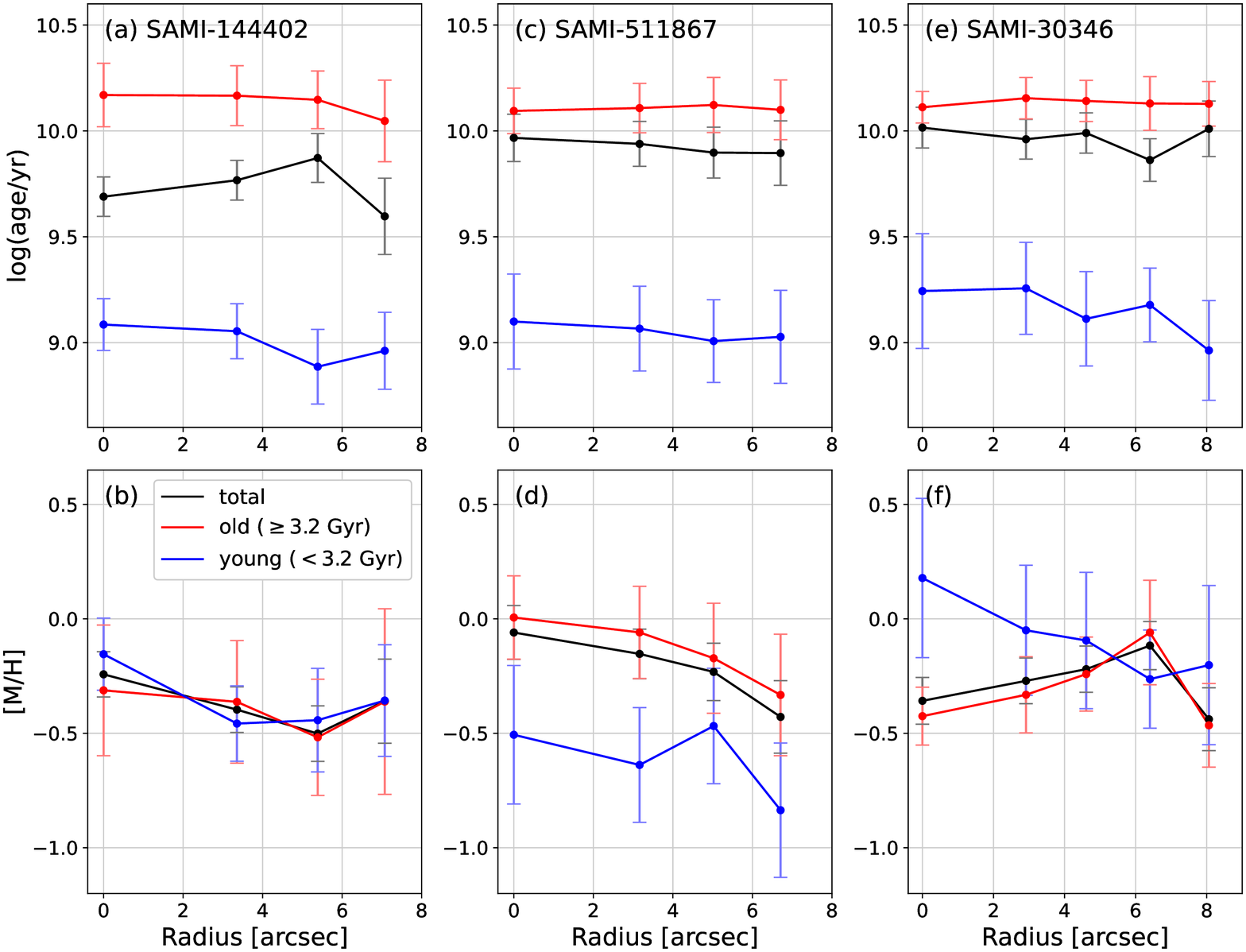}
\caption{(a) Mean stellar age profiles of SAMI-144402, for the total mean stellar populations (black line), for young components (blue line), and for old components (red lines). The errorbar was calculated using the S/N and $f_{lum}$ of each bin, based on Figure~\ref{age2}. (b) Mean stellar metallicity profiles of SAMI-144402, and each errorbar is based on Figure~\ref{met2}. (c) - (d) The same as (a) - (b), but for SAMI-511867. (e) - (f) The same as (a) - (b), but for SAMI-30346. \label{examprof}}
\end{figure*}

Finally, we present some examples of how to apply our results to the actual analysis of observed galaxy spectra. Figures~\ref{exam1} - \ref{exam3} show the maps of several parameters including total and age-divided mean stellar populations, for three SAMI galaxies. To secure sufficient S/N necessary for high reliability, annularly-binned cubes from the final data release of the SAMI Galaxy Survey are used. The radial profiles of age and metallicity for the three galaxies are compared in Figure~\ref{examprof}.
Following the test results in Section~\ref{emex}, we assume that the impact of dust attenuation and emission lines on ADP estimation may be insignificant.

SAMI-144402 (Figure~\ref{exam1}) is a merger remnant galaxy with large tidal arms. It is noted that the age profile of the total mean stellar population of this galaxy shows a positive gradient (i.e., older at outskirts) out to $R\sim 6''$ (Figure~\ref{examprof}(a)). However, when the stellar populations are divided into young and old components, neither of the two components shows a positive age gradient. Thus, if we trust the estimated ADPs, the positive age gradient of this galaxy may be because the young stars are not evenly distributed along radius but concentrated in the center. On the other hand, the metallicity profiles of the young and old stellar components agree well with that of the total mean stellar populations, showing a negative gradient (Figure~\ref{examprof}(b)). Thus, the formation source of the young stars in SAMI-144402 may be the in-situ cold gas of this galaxy, rather than gas accretion from external sources. As a possible scenario, when we consider the morphology of this galaxy, an intruder might have stimulated the in-situ gas of the host galaxy, triggering new star formation at the center.

SAMI-511867 (Figure~\ref{exam2}) is a spiral galaxy with the spiral arms forming a pseudo-ring. Although the mass fraction of its young stellar component is not high ($\lesssim 20$ per cent), its luminosity fraction is pretty high ($\gtrsim 60$ per cent), which makes the uncertainty of young stellar components not too large. There are two notable features in this galaxy: (1) the metallicity of young stellar components is obviously lower than that of old stellar components, and (2) the young stars in the pseudo-ring ($R \sim 5''$) appears to have slightly enhanced metallicity, compared to the young stars in other parts. However, since the amplitude of the enhancement is not large, compared to the uncertainty interpolated from Figure~\ref{met2}, the metallicity-enhanced ring in this galaxy remains as a weak possibility here. Nevertheless, the systematically lower metallicity of the young stellar component compared to that of the old stellar component implies that the recent star formation activity all over the radius in SAMI-511867 may have originated from metal-poor gas, possibly injected from outside.

SAMI-30346 (Figure~\ref{exam3}) is a spiral galaxy with flocculent spiral arms. The total mean stellar population of this galaxy shows an unusual radial profile: the metallicity increases along radius out to $R\sim 6''$. Since usual massive galaxies tend to have negative gradients in their metallicity profiles \citep[e.g.,][]{str76,sha83,pas14}, this reversed gradient implies that this galaxy may have a unique assembly history (e.g., recent major merger). When the populations are divided into young and old components, another intriguing feature is found that the metallicity profile of the young stellar component have a negative gradient unlike the old stellar component. Moreover, the metallicity of the young stellar component appears to be larger than that of the old stellar component in the galaxy center. One possible scenario for these features is that this galaxy may have experienced a major merger event, and as a result, the old stars in this galaxy may have spatially scattered, forming an unusually positive gradient of metallicity. Right after that, the remnant gas of the major merger, which has already been enriched to some extent and temporarily scattered around the galaxy, may have fallen into the galaxy again, and forming new stars with further metal enrichment particularly at the centre.

\section{CONCLUSIONS}\label{conclude}

We introduced a new methodology using the age-divided mean stellar populations (ADPs) estimated from the full spectrum fitting, as a practical tracer of the star formation and chemical evolution history of a galaxy. To estimate the statistical reliability of the mass-weighted total mean stellar populations and ADPs, we generate 10,000 artificial galaxy spectra, each of which consists of five stellar population components with random ages, metallicities, and mass fractions, to which we added realistic (imitating the SAMI survey data) noise as a function of wavelength. We reconstruct the template composition of each galaxy spectrum by using the pPXF package.
The $\Delta$ (output $-$ input) in the total mean age and metallicity appear to significantly depend on S/N, while there are only small differences among several pPXF options. The effect of the age-metallicity degeneracy is not serious unless S/N is too small (e.g. $< 30$).

In the separate estimation of young ($< 10^{9.5}$ yr) and old ($\geq 10^{9.5}$ yr) stellar components, $\Delta$log(age/yr) and $\Delta$[M/H] appear to significantly depend on luminosity fraction of each component as well as the total S/N. The degeneracy among age, metallicity, and mass fraction appears to be mostly weak ($|\mathcal{R}| \lesssim 0.3$). However, $\Delta$log(age/yr) of young stellar components are strongly correlated with their $\Delta f_{\textrm{mass}}$ ($\mathcal{R} \sim 0.6$) regardless of S/N, which is interpreted to be a natural result from the similarity between very young stars with a small fraction and moderately young stars with a moderate fraction, when mixed with an old stellar component. Despite several caveats such as the partial degeneracy and the intrinsic limits of this work (e.g., gas and dust were not fully considered), the overall reliability of ADPs appears to be mainly determined by total S/N and luminosity fractions of young and old components. These results (or the methodology) can be applied to the analysis of the star formation and chemical evolution history from an observed galaxy spectrum with high S/N, in a relatively simple way but with caution. More systematic studies by applying this methodology to a larger sample of the SAMI galaxies will come soon.

\section*{Acknowledgements}

We thank the anonymous referee for the constructive suggestions and helpful comments.
This work was supported by the Korea Astronomy and Space Science Institute under the R{\&}D program (Project No.~2023-1-830-01) supervised by the Ministry of Science and ICT (MSIT).
JHL, MP, HJ, and SO acknowledge support from the National Research Foundation of Korea (NRF) grant funded by the Korea government(MSIT) (No.~2022R1A2C1004025, No.~2022R1C1C2006540, No.~2019R1F1A1041086, and No.~2020R1A2C3003769, respectively). This research was partially supported by the Australian Research Council Centre of Excellence for All Sky Astrophysics in 3 Dimensions (ASTRO 3D), through project number CE170100013.

\section*{Data Availability}

The tables summarising the key results are available as online material.





\bsp	
\label{lastpage}
\end{document}